\documentclass[aps,prd,onecolumn,groupedaddress,showpacs,nofootinbib,amssymb,eqsecnum]{revtex4}
\usepackage{amsmath}
\usepackage{amssymb}
\usepackage{amsfonts}
\usepackage{graphicx,bm}
\usepackage{dcolumn}
\usepackage{epsfig}
\usepackage{color,amsxtra}
\usepackage{epsf}
\usepackage{enumerate}
\usepackage{hhline}
\usepackage{array}
\usepackage{tabularx}
\newcommand{\be}{\begin{equation}}
\newcommand{\ee}{\end{equation}}
\newcommand{\bea}{\begin{eqnarray}}
\newcommand{\eea}{\end{eqnarray}}
\newcommand{\beaa}{\begin{eqnarray*}}
\newcommand{\eeaa}{\end{eqnarray*}}

\newcommand{\nn}{\nonumber \\}
\newcommand{\e}{\mathrm{e}}
\newcommand{\tr}{\mathrm{tr}\,} 
\allowdisplaybreaks[4]

\newcommand{\Eqn}[1]{&\hspace{-0.2em}#1\hspace{-0.2em}&}

\def\be{\begin{equation}}
\def\ee{\end{equation}}
\def\bea{\begin{eqnarray}}
\def\eea{\end{eqnarray}}
\def\tr{\mathrm{tr}\, }
\def\nn{\nonumber \\}
\def\e{\mathrm{e}}
\begin{document}
%\draft

%\tolerance=5000

%%%%%%%%%%%%%%%%%%%%%
%  Title
%%%%%%%%%%%%%%%%%%%%%
\title{Bounce cosmology from $F(R)$ gravity and $F(R)$ bigravity}
\author{Kazuharu Bamba$^{1,}$\footnote{
E-mail address: bamba@kmi.nagoya-u.ac.jp}, 
Andrey N. Makarenko$^{2,}$\footnote{
E-mail address: andre@tspu.edu.ru}, \\ 
Alexandr N. Myagky$^{3,}$\footnote{
E-mail address: alex7604@mail.ru}, 
Shin'ichi Nojiri$^{1, 4,}$\footnote{E-mail address:
nojiri@phys.nagoya-u.ac.jp} 
and 
Sergei D. Odintsov$^{2, 5, 6}$\footnote{
E-mail address: odintsov@ieec.uab.es}
}
\affiliation{
$^1$Kobayashi-Maskawa Institute for the Origin of Particles and the
Universe,
Nagoya University, Nagoya 464-8602, Japan\\ 
$^2$ Tomsk State Pedagogical University, Kievskaya Avenue, 60, 
634061, Tomsk, Russia\\
$^3$ Tomsk Polytechnic University, Lenin Avenue, 30, 
634050, Tomsk, Russia\\
$^4$Department of Physics, Nagoya University, Nagoya 464-8602, Japan\\
$^5$Instituci\`{o} Catalana de Recerca i Estudis Avan\c{c}ats (ICREA), 
Barcelona, Spain\\
$^6$Institut de Ciencies de l'Espai (CSIC-IEEC), 
Campus UAB, Facultat de Ciencies, Torre C5-Par-2a pl, E-08193 Bellaterra
(Barcelona), Spain
}

%\date{\today}

%%%%%%%%%%%%%%%%%%%%%
%  Abstract
%%%%%%%%%%%%%%%%%%%%%
\begin{abstract}

We reconstruct $F(R)$ gravity models with exponential and power-law forms of the scale factor in which bounce cosmology can be realized. 
We explore the stability of the reconstructed models with 
analyzing the perturbations from the background solutions. 
Furthermore, we study an $F(R)$ gravity model with a sum of exponentials 
form of the scale factor, where the bounce in the early universe as well as 
the late-time cosmic acceleration can be realized in a unified manner. 
As a result, we build a second order polynomial type model in terms of $R$ 
and show that it could be stable. 
Moreover, when the scale factor is expressed by an exponential form, 
we derive $F(R)$ gravity models of a polynomial type 
in case of the non-zero spatial curvature 
and that of a generic type in that of the zero spatial curvature. 
In addition, for an exponential form of the scale factor, 
an $F(R)$ bigravity model realizing the bouncing behavior 
is reconstructed. It is found that in both the physical and reference metrics 
the bouncing phenomenon can occur, although in general 
the contraction and expansion rates are different each other. 

\end{abstract}
%%%%%%%%%%%%%%%%%%%%%

%----------------------------
%\pacs{
%Keywords:
%}
%\preprint{}
%\hspace{13.0cm}
%----------------------------

\maketitle
%==============================================================================

%%%%%%%%%%%%%%%%%%%%%%%%%%%
%%%  Sec. I
%%%%%%%%%%%%%%%%%%%%%%%%%%%
\section{Introduction}

According to recent cosmological observations in terms of 
Supernovae Ia~\cite{SN1}, large scale structure~\cite{LSS} 
with the baryon acoustic oscillations~\cite{Eisenstein:2005su}, 
cosmic microwave background radiation~\cite{WMAP}, 
and weak lensing~\cite{Jain:2003tba}, 
the current expansion of the universe is accelerating. 
We suppose that the universe is homogeneous, as suggested by observations. 
We have two representative procedures to explain the cosmic 
acceleration at the present time. 
One is the introduction of the so called dark energy with 
negative pressure in general relativity 
(for reviews on dark energy, see, e.g.,~\cite{R-DE}). 
The other is the modification of gravity on the large distances. 
As a simple way of modification of gravity, 
here we concentrate on $F(R)$ gravity~\cite{Capozziello:2002rd, Carroll:2003wy, Nojiri:2003ft} 
(for reviews, see, for example,~\cite{R-F(R)-NO-CF-CD-DS}). 

On the other hand, as a cosmological scenario in the early universe, 
there exists the so-called matter bounce scenario~\cite{B-R-BCS}, 
in which (i) in the initial phase of the contraction, the universe is 
at the matter-dominated stage, 
(ii) there happens a bounce without any singularity, 
and (iii) the primordial curvature perturbations with the observed spectrum 
can also be generated (for a review on bounce cosmology, 
see~\cite{Novello:2008ra}). 
It is known that in this scenario, 
there is the BKL instability~\cite{Belinsky:1970ew} 
leading to an anisotropic universe after the contracting phase. 
%%%%%%%%
In the framework of the Ekpyrotic scenario~\cite{Khoury:2001wf}, 
the resolution of such an instability to produce the anisotropy 
of the universe~\cite{Erickson:2003zm} and 
intrinsic problems in the bouncing process~\cite{X-S} 
have been studied in Refs.~\cite{Cai:2012va, Cai:2013vm}. 
Recently, the curvature perturbations generated in the matter bounce cosmology with two fields was re-examined in more detail in Ref.~\cite{Cai:2013kja}. 
Furthermore, as recent related studies, 
cyclic cosmology~\cite{Bars:2013vba}, cosmological 
perturbations in bounce cosmology without singularities~\cite{Xue:2013bva, QGS-QECLZ-Q-Q}, 
and properties of cosmological perturbations around the bouncing epoch~\cite{Pinto-Neto:2013npa, LGP-PFZ} have been investigated. 
%%%%%%%%
Here, it should be noted that when there is a massive scalar field, 
the scale factor and the Riemann curvature can have the bouncing behaviors 
with the positive spatial curvature $k (>0)$, 
which will be presented in Sec.~VI. 
This was first shown in Ref.~\cite{P-S-BS}. 
In addition, for the Starobinsky model proposed 
in Ref.~\cite{Starobinsky:1980te}, 
there exists a solution with the non-zero spatial curvature $k (\neq 0)$ 
in which the scale factor behaves a bounce. 
%%%%%
Moreover, in Refs.~\cite{Garriga:2013cix, Gupt:2013poa, Piao-cycle} 
it has been investigated that 
with a simply modified Friedmann equation, 
the bouncing behavior of the scale factor would occur 
at the time when the energy density of matter evolves into a critical value. 
Thus, it has been shown that 
big crunch singularities 
of negative-energy (Anti-de Sitter) bubbles 
in the multiverse can be removed. 
%%%%%
Also, the behavior of bounce in anisotropic cosmology 
in $F(R)$ gravity~\cite{Leon:2013bra} 
and a bounce in modified gravty theories~\cite{BouhmadiLopez:2012qp} 
have recently been discussed. 
We further mention that the form of $F(R)$ leading to non-singular 
bounce cosmologies has been derived in Ref.~\cite{Olmo:2008nf}. 
%%%%%
%%%
Furthermore, bounces in gravity inspired by string theories~\cite{Biswas:2005qr} and non-local gravities~\cite{BKMV-BKM} have been studied.
%%%

%%%%%
In addition, it has recently been revealed 
that a massive graviton can lead to the current cosmic acceleration. 
At the early stage, 
the Fierz-Pauli (FP) action~\cite{F-P} was considered to describe 
a linearized or free theory of massive gravity 
(for reviews, see, for example,~\cite{Hinterbichler:2011tt, 
Goldhaber:2008xy}). 
Recently, the de Rham, Gabadadze, Tolley (dRGT) 
theory~\cite{deRham:2010ik, deRham:2010kj} 
and the Hassan-Rosen (HR) theory~\cite{H-R}, 
which are non-linear massive gravity theories, 
have been proposed. 
These theories have two desirable properties: One is there is not 
the Boulware-Deser (BD) ghost~\cite{BD-G, Boulware:1974sr}. 
Another is in the massless limit of the mass of massive graviton 
the van Dam-Veltman-Zakharov (vDVZ) discontinuity~\cite{vDVZ-D} 
can be screened through the Vainstein mechanism~\cite{Vainshtein:1972sx}. 
The latter is a similar feature appearing in the Galileon models~\cite{G-M} 
due to the operation on the Dvali-Gabadadze-Porrati (DGP) brane world 
scenarios~\cite{DGP-D}. 
Currently, in various aspects, 
massive gravity and bi-metric gravity have extensively been 
studied in the literature~\cite{Hassan:2011tf, Hassan:2011vm, Golovnev:2011aa, NMG-K, Hassan:2012qv, Hassan:2011ea, K-N-T, D'Amico:2011jj, Hinterbichler:2012cn, Baccetti:2012bk, Kobayashi:2012fz, Nomura:2012xr, Saridakis:2012jy, ZSS-SYZ, Mohseni:2012ug, Damour:2002wu, Volkov:2011an, vonStrauss:2011mq, Berg:2012kn, Akrami:2012vf, BezerradeMello:2012nq, deRham:2011pt, Guarato:2013gba}.

However, thanks to recent works~\cite{no-f-FLRW}, 
it has been found that in such non-linear massive gravity theories, 
the flat homogeneous and isotropic 
Friedmann-Lema\^{i}tre-Robertson-Walker (FLRW) universe, 
which is supported by various cosmological observations, 
cannot be stable. 
Hence, massive gravity theories in the context of general relativity 
explained above, which is called ``the massive general relativity (GR)'' 
in the literature, 
have been extended, for instance, 
an extended version of the dRGT theory~\cite{Hinterbichler:2013dv},  
a massive bi-metric $F(R)$ gravity theory~\cite{Nojiri:2012zu, Nojiri:2012re}, 
a new massive $F(R)$ gravity~\cite{Kluson:2013yaa} proposed very recently, 
a scale invariant theory with a dilaton field, or 
which is called the ``Quasi-Dilaton'' massive gravity 
(QMG)~\cite{D'Amico:2012zv, Gannouji:2013rwa} 
and its extended versions~\cite{Kluson:2013jea}, 
and mass varying scenario in which a massive graviton mass depends on 
a dynamical scalar field~\cite{MVMG}. 
%%%%%

In this paper, with the procedure proposed in Ref.~\cite{Nojiri:2009kx}, 
which corresponds to a kind of simpler and more useful reconstruction method 
made by developing the formulation in Ref.~\cite{Capozziello:2005tf}, 
we derive $F(R)$ gravity models in which bounce cosmology can occur. 
In particular, in the flat FLRW universe 
we perform the analysis for two cases 
that the scale factor is described by exponential and power-law forms. 
We study the perturbations from the background solutions and 
explicitly explore the stability conditions for these models to be stable. 
%%%
In addition, we investigate an $F(R)$ gravity model with 
the scale factor having a sum of exponentials form, where 
the unification of the bouncing behavior in the early universe and 
the late-time cosmic acceleration can be realized.  
%%%
Furthermore, 
in the FLRW universe with non-zero spatial curvature, 
for an exponential form of the scale factor, 
we reconstruct $F(R)$ gravity models in which 
a function of $F(R)$ is expressed by a polynomial in terms of $R$. 
Also, for the scale factor with an exponential form, 
in the flat FLRW universe, we build $F(R)$ gravity models 
by using the reconstruction method~\cite{odin1} and 
explore the stability conditions. 
%%%
We also reconstruct an $F(R)$ bigravity model realizing bounce cosmology. 
Incidentally, the bouncing behavior and cyclic cosmology in extended 
non-linear massive gravity~\cite{Cai:2012ag} and 
bounce cosmology in bigravity~\cite{Capozziello:2012re} 
have been investigated. 
%%%

%%%%%%%%
Here, we clarify our purpose of this study. 
As the first step, in this work 
we reconstruct $F(R)$ gravity and $F(R)$ bigravity models 
with the bouncing behavior. 
In particular, for $F(R)$ gravity, 
we build models in which not only the bounce in the early universe 
but also the late-time cosmic acceleration occurs 
and examine the stability of these models. 
At the current stage, these models are still toy models. 
However, we note that some of the considered models with an $R^2$ term are 
known to be viable models for the early-time inflation. 
Moreover, it should be emphasized that bounce cosmology may be a natural part of the complete and viable history of the universe. 
This is the reason to study better such cosmologies. 
Our final goal is to construct the so called viable 
$F(R)$ gravity and $F(R)$ gravity models, in which 
all the cosmological various processes of expansion history of the universe 
with a bounce can be realized. 
This developed subject should be executed as another separate work 
in the near future. 
%%%%%%%%
%%%%%%%%%%%%
%%% Unit %%%
We use units of $k_\mathrm{B} = c_{\mathrm{l}} = \hbar = 1$, 
where $c_{\mathrm{l}}$ is the speed of light, and denote the
gravitational constant $8 \pi G_{\mathrm{N}}$ by
${\kappa}^2 \equiv 8\pi/{M_{\mathrm{Pl}}}^2$
with the Planck mass of $M_{\mathrm{Pl}} = G_{\mathrm{N}}^{-1/2} = 1.2 \times 
10^{19}$\,\,GeV. 
%%%%%%%%%%%%

%%%%% Structure %%%%%
The paper is organized as follows. 
In Sec.\ II, we explain a reconstruction method of $F(R)$ gravity. 
With this procedure, 
we derive $F(R)$ gravity models realizing bounce cosmology 
in Sec.\ III. 
Furthermore, in Sec.\ IV we examine the stability of 
the reconstructed $F(R)$ gravity models. 
In Sec.\ V, we also build an 
$F(R)$ model where both the bounce in the early universe and 
the late-time cosmic acceleration can occur in a unified manner. 
In Sec.\ VI, we investigate an exponential form of the scale factor 
for the non-zero spatial curvature, 
while in Sec.\ VII, we explore it for the zero spatial curvature. 
Moreover, in Sec.\ VIII we reconstruct $F(R)$ bigravity models in which 
the bouncing phenomenon can happen. 
In Sec.\ IX, conclusions are presented. 
%%%%%%%%%%%%%%%%%%%%%

%%%%%%%%%%%%%%%%%%%%%%%%%%%
%%%  Sec. II
%%%%%%%%%%%%%%%%%%%%%%%%%%%
\section{Reconstruction method of $F(R)$ gravity}

In this section, we explain 
the reconstruction method of $F(R)$ gravity~\cite{Nojiri:2009kx}. 
The action of $F(R)$ gravity with matter is expressed as 
\begin{equation} 
S = \int d^4 x \sqrt{-g} 
\frac{F(R)}{2\kappa^2} + 
\int d^4 x 
{\mathcal{L}}_{\mathrm{M}} \left( g_{\mu\nu}, {\Psi}_{\mathrm{M}} \right)\,,
\label{eq:2.1}
\end{equation}
with ${\mathcal{L}}_{\mathrm{M}}$ the matter Lagrangian 
and ${\Psi}_{\mathrm{M}}$ matter fields. 

In the flat Friedmann-Lema\^{i}tre-Robertson-Walker (FLRW) 
universe, the metric is given by 
\begin{equation} 
ds^2 = -dt^2 + a^2(t) \sum_{i=1,2,3}\left(dx^i\right)^2\,, 
\label{eq:FR7-3-Add-01}
\end{equation}
with $a$ the scale factor. 

Here, we introduce the number of $e$-folds defined by 
$N \equiv \ln \left(a/a_* \right)$, where $a$ is a scale factor and 
$a_*$ is a value of $a$ at a time $t_*$. 
When we take $t_* = t_0$ with $t_0$ the present time and 
and $a_* = a_0$ at $t= t_0$, 
we can also define the redshift z as $z \equiv a_0/a - 1$. 
Moreover, the Hubble parameter is given by 
$H \equiv \dot{a}/a$, where the dot denotes the time derivative of 
$\partial/\partial t$, and we describe it 
by using a function of $\tilde{g} (N)$ as 
$H=\tilde{g} (N = -\ln \left( 1+z \right))$. 
Furthermore, we write $H^2$ as 
$H^2 \equiv G (N) = \tilde{g}^2 (N)$ 
with $G (N)$ a function of $N$. 
%%%%%%%%
With the quantities defined above, in this background 
the Friedmann equation reads
\begin{equation}
9G (N(R)) \left( 4G'(N(R)) + 
G''(N(R)) \right) \frac{d^2 F(R)}{d R^2} 
-3\left( G(N(R)) + \frac{1}{2} G'(N(R)) \right) 
\frac{d F(R)}{d R} 
+\frac{1}{2}F(R) -\kappa^2 \rho_{\mathrm{M}} = 0\,,
\label{eq:2.2}
\end{equation}
with 
\begin{equation} 
\rho_{\mathrm{M}} = \sum_{i} \rho_{\mathrm{M} \, i 0} 
a^{-3\left(1+w_i \right)} 
= \sum_{i} \rho_{\mathrm{M} \, i 0} a_0^{-3\left(1+w_i \right)} 
\exp \left[-3 \left(1+w_i \right) N \right]\,. 
\label{eq:2.3}
\end{equation}
Here, $\rho_{\mathrm{M}}$ is the sum of energy density of 
all matters assumed to be fluids with a constant equation of state $w_i$ 
defined as $w_i \equiv P_{\mathrm{M} \, i}/ \rho_{\mathrm{M} \, i}$, 
where the subscription ``$i$'' shows the label of the fluids 
and $\rho_{\mathrm{M} \, i}$ and $P_{\mathrm{M} \, i}$ are 
the energy density and pressure of the $i$-th fluid, respectively, 
$\rho_{\mathrm{M} \, i 0}$ is a constant, 
and the prime denotes the derivative with respect to $N$ as 
$G'(N) \equiv dG/dN$ and $G''(N) \equiv d^2G/dN^2$.

%%%%%%%%%%%%%%%%%%%%%%%%%%%
%%%  Sec. III
%%%%%%%%%%%%%%%%%%%%%%%%%%%
\section{$F(R)$ gravity realizing bounce cosmology}

In this section, we study  
the cosmological background evolutions in the 
matter bounce cosmology and 
reconstruct $F(R)$ gravity models realizing it. 

%%%%%%%%%%%%%%%%%%%%%%%%%%%
%%%  Sec. III A
%%%%%%%%%%%%%%%%%%%%%%%%%%%
\subsection{Exponential model} 

We examine the case that the scale factor is expressed 
by an exponential form. 
For instance, we consider a bouncing solution which behaves as
\be
\label{b1}
a(t) \sim \e^{\alpha t^2}\, .
\ee
Here, $\alpha$ is a constant with the dimension of mass squared 
($[\mathrm{Mass}]^2$). 
%%%
In the following, we set $N \equiv \ln a (t) /a (t=0)$, where 
$a (t=0) =1$ because we study the bouncing behavior around 
$t=0$. 
%%%
We now use the reconstruction in Ref.~\cite{Nojiri:2009kx}. {}From 
Eq.~(\ref{eq:2.2}), 
we solve the following differential equation: 
\bea
\label{b2}
&&
0 = - 9 G \left( N \left( R \right) \right) \left( 4 G' \left( N \left( R \right) \right) 
+ G'' \left( N \left( R \right) \right) \right) \frac{d^2 F(R)}{dR^2} 
\nonumber \\
&&
+ \left( 3 G \left( N \left( R \right) \right) + \frac{3}{2} 
G'  \left( N \left( R \right) \right) \right) \frac{dF(R)}{dR} - 
\frac{F(R)}{2}\, . 
\eea
Here, we have neglected a contribution from matters and 
$G(N)=H(N)^2$ and the scalar curvature $R$ is given by 
\be
\label{b3}
R = 3 G' (N) + 12 G (N) \, .
\ee 
For the model (\ref{b1}), we find
\be
\label{b4}
N = \alpha t^2\, ,\quad H = \dot N = 2 \alpha t\, ,
\ee
which give
\be
\label{b5}
G(N) = 4 \alpha N \, ,\quad R= 12 \alpha \left( 1 + 4 N \right)\, ,
\ee
and therefore
\be
\label{b6}
N = - \frac{1}{4} + \frac{R}{48 \alpha}\, .
\ee
Then, Eq.~(\ref{b2}) has the following form:
\be
\label{b7}
0 = - 144 \alpha^2 \left( - 1 + \frac{R}{12\alpha} \right) \frac{d^2 F}{dR^2} 
+ 3\alpha \left( 1 + \frac{R}{12\alpha} \right) \frac{dF}{dR} - \frac{F}{2}\, .
\ee
A solution of (\ref{b7}) is given by
\be
\label{b8}
F(R) = \frac{1}{\alpha}R^2 - 72R + 144 \alpha\, .
\ee
%%%
%We remark that a sort of this model also realizes 
%the Starobinsky inflation~\cite{Starobinsky:1980te}, 
%and therefore that it is a realistic model to describe 
%the early universe. 
%%%

\begin{center}
\begin{figure}[tbp]
\resizebox{!}{6.5cm}{
   \includegraphics{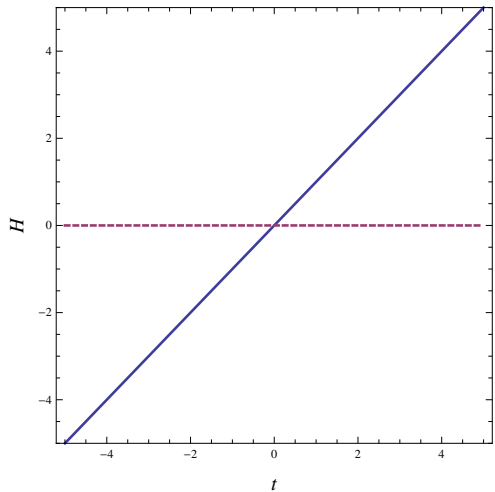}
                  }
\caption{The Hubble parameter (solid line) in the second 
relation in (\ref{b4}) for $\alpha = 1/2$ 
around a bounce at $t=0$. 
The dotted line shows $H=0$. 
} 
\label{FR7-fig-1}
\end{figure}
\end{center}

%%%%%%%%
In Fig.~\ref{FR7-fig-1}, 
we show the behavior of the Hubble parameter in the second 
relation in (\ref{b4}) for $\alpha = 1/2$ 
around a bounce at $t=0$. {}From this figure, we see that 
before the bounce ($t<0$), $H <0$, while after it ($t>0$), 
$H>0$. Thus, the bouncing behavior occurs. 
%%%%%%%%

%%%%%%%%%%%%%
%%% Fig-1 %%%
%%%%%%%%%%%%%%%%%%%%%%%%%%%%%%%%%%%%%%%%%%%%%%%%%%%%%%
%%%%%%%%%%%%%%%%%%%%%%%%%%%%%%%%%%%%%%%%%%%%%%%%%%%%%%

%%%%%%%%
It should clearly be mentioned that the metric 
with the scale factor (\ref{b1}) 
does not have a finite maximum in the Riemann curvature, 
whereas the Riemann curvature takes its minimum by modulus 
in the bounce epoch. 
Thus, this space-time is irrelevant to the thing necessary to 
remove a cosmological singularity. 
By the same reason as written above, in the model described by Eq.~(\ref{b8}) 
the Starobinsky inflation~\cite{Starobinsky:1980te} cannot be realized. 
When the Starobinsky inflation occurs, the scale factor at the slow-roll 
inflationary stage is given by 
$a(t) \propto \exp \left(H_1 t - M^2 t^2/12 \right)$. 
%%%%%%%%

%%%%%%%%
Also, we explore the stability with respect to tensor perturbations, 
namely, the required condition $F'(R) > 0$. 
%%%%%
It follows from the second relation in (\ref{b5}) with the first one 
in (\ref{b4}) and Eq.~(\ref{b8}) that we have 
$F'(R) = \left(2/\alpha\right) \left(R-36\alpha\right) 
= 48 \left(2 \alpha t^2 -1 \right)$. 
Hence, when a bounce occurs at $t =0$, we find $F'(R) <0$. 
We also see that $F'(R) = 0$ at $R =36 \alpha$. 
As a result, the bounce of the scale factor in Eq.~(\ref{b1}) occurs 
in the unphysical regime of a negative effective 
gravitational constant, so that graviton can become a ghost. 
In addition, for $\alpha>0$, $F''(R) =2/\alpha > 0$, and therefore 
the stability condition for the cosmological 
perturbations~\cite{Nojiri:2003ft, DK-F-SHS} can be satisfied. 
%%%
Moreover, in Refs.~\cite{Nariai:1973eg} and \cite{Gurovich:1979xg} 
it has been found that even at the classical level, 
it is not able to pass the point in which $F'(R) = 0$ for a finite $R$ 
because in a generic solution a strong anisotropic curvature singularity 
appears. 
%%%
%%%%%%%%

%%%%%%%%%%%%%%%%%%%%%%%%%%%
%%%  Sec. III B
%%%%%%%%%%%%%%%%%%%%%%%%%%%
\subsection{Power-law model} 

On the other hand, it is known that for 
the case in which the scale factor is expressed 
by a power-law model, given by 
\begin{equation} 
a(t) = \bar{a} \left( \frac{t}{\bar{t}} \right)^q + 1\,,  
\label{eq:3.9}
\end{equation}
where $\bar{a} (\neq 0)$ a constant, $\bar{t}$ is a fiducial time, and $q = 2n$ with $n$ is an integer, 
a power-law model of $F(R)$ gravity would be reconstructed. 
In this case, we acquire 
\begin{eqnarray}
N \Eqn{=} \ln \left[ \bar{a} \left( \frac{t}{\bar{t}} \right)^q +1 \right]\,, 
\label{eq:3.10} \\
H \Eqn{=} \frac{\bar{a} q \left(1/\bar{t}\right) \left(t/\bar{t}\right)^{q-1} }{\bar{a} \left(t/\bar{t}\right)^{q} +1}
\label{eq:3.11} 
\end{eqnarray}
With Eqs.~(\ref{b3}), (\ref{eq:3.10}), (\ref{eq:3.11}) and $G=H^2$, 
we find 
\begin{eqnarray}
G(N) \Eqn{=} \left(\frac{q}{\bar{t}}\right)^2 \bar{a}^{2/q} \e^{-2N} \left(\e^{N}-1\right)^{2\left(1-1/q\right)}\,,  
\label{eq:3.13} \\
R \Eqn{=} 6 \left(\frac{q}{\bar{t}}\right)^2 \bar{a}^{2/q} 
\e^{-N} \left(\e^{N}-1\right)^{1-2/q} \left(2 -\frac{1}{q} \right)\,. 
\label{eq:3.14} 
\end{eqnarray}
Around the bounce behavior, we have $N \simeq 0$. 
Hence, by adopting an approximation $\e^{N} \simeq 1$ to Eq.~(\ref{eq:3.14}), we obtain 
$R \simeq 6 \left(q/\bar{t}\right)^2 \bar{a}^{2/q} 
\left(\e^{N}-1\right)^{1-2/q} \left(2 -1/q \right)$. 
With this approximate expression of $R$, Eq.~(\ref{b2}) reads
\begin{equation}
-\frac{q-2}{2q-1} R^2 \frac{d^2 F(R)}{dR^2} 
+R \frac{dF(R)}{dR} -F(R) = 0\,, 
\label{eq:3.16} 
\end{equation}
where we have also neglected the matter contributions. 
As a solution, we have
\begin{eqnarray}
&&
F(R) = \bar{F} R^{\beta}\,,
\label{eq:3.17} \\
&&
\beta = 1\,, 
%\quad \mathrm{or} 
\quad \frac{2q-1}{q-2}\,,
\end{eqnarray}
with $\bar{F} (\neq 0)$ a constant. 
%%%%%%%%
It has first been shown in Ref.~\cite{Muller:1989rp} that 
there exist power-law solutions for such a monomial form of $F(R)$. 
%%%%%%%%

\begin{center}
\begin{figure}[tbp]
\resizebox{!}{6.5cm}{
   \includegraphics{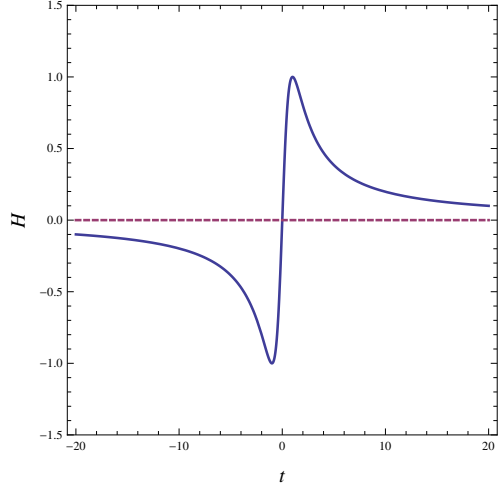}
                  }
\caption{The Hubble parameter in Eq.~(\ref{eq:3.11}) 
around a bounce at $t=0$ for $\bar{a} =1.0$, $q=2$ with $n=1$, 
and $\bar{t}=1$. 
Legend is the same as Fig.~\ref{FR7-fig-1}. 
} 
\label{FR7-fig-2}
\end{figure}
\end{center}

%%%%%%%%
In Fig.~\ref{FR7-fig-2}, for $\bar{a} =1.0$, $q=2$ with $n=1$, 
and $\bar{t}=1$, 
we depict the behavior of the Hubble parameter in Eq.~(\ref{eq:3.11}) 
around a bounce at $t=0$. It follows from this figure that 
before the bounce ($t<0$), $H <0$, 
whereas after it ($t>0$), $H>0$, similarly to that 
in Fig.~\ref{FR7-fig-1}. As a result, the bouncing behavior happens. 
%%%%%%%%

%%%%%%%%%%%%%
%%% Fig-2 %%%
%%%%%%%%%%%%%%%%%%%%%%%%%%%%%%%%%%%%%%%%%%%%%%%%%%%%%%
%%%%%%%%%%%%%%%%%%%%%%%%%%%%%%%%%%%%%%%%%%%%%%%%%%%%%%

%%%%%%%%
%%%%%%%%
%%%%%%%%
For the scale factor in Eq.~(\ref{eq:3.9}), from Eq.~(\ref{eq:3.17}) 
we obtain $F'(R) = \bar{F} \beta R^{\beta-1}$, where $R$ 
is given by Eq.~(\ref{eq:3.14}). 
When a bounce happens, we have $N \simeq 0$ and thus 
$R \geq 0$. As a consequence, we see that for $\bar{F} >0$, $F'(R) >0$. 
Furthermore, for $\beta>1$, i.e., $q=2n$ with $n>1$, 
$F''(R) = \bar{F} \beta \left(\beta-1\right) R^{\beta-2} >0$. 
Hence, the condition of the stable cosmological 
perturbations can be met~\cite{Nojiri:2003ft, DK-F-SHS}. 
We mention that in this power-law model in Eq.~(\ref{eq:3.17}) with 
$\bar{F} >0$, in the limit $R \to 0$, namely, in the bounce, 
we find $F'(R) = 0$ at $R =0$. In this case, since $R$ vanishes when 
$F'(R) =0$, $F'(R)$ does not pass the point where $F'(R) =0$~\cite{Nariai:1973eg, Gurovich:1979xg}. 
%%%%%%%%
%%%%%%%%
%%%%%%%%

We also remark that 
in the matter bounce cosmology with two fields~\cite{Cai:2013kja}, 
for $a \propto \left( t-\bar{t} \right)^{s}$, 
cosmological background evolutions consist of the following 
four phases: 
(i) matter contraction phase, 
(ii) the Ekpyrotic contraction phase, 
(iii) bounce phase, 
and (iv) fast-roll expansion phase. 
In the matter contraction, the Ekpyrotic contraction, and fast-roll expansion 
phases, the scale factor $a$ behaves as power-law type in 
Eq.~(\ref{eq:3.9}), 
while in the bounce phase, $a$ evolves as exponential type in 
Eq.~(\ref{b1}). 
For the matter contraction phase, we find $s = 2/3$, 
for the Ekpyrotic contraction phase, $s$ would not be set to 
a specific value, 
whereas in the fast-roll expansion phase, we have $s=1/3$. 
On the other hand, for the Ekpyrotic contraction phase, 
if the scale factor is described by Eq.~(\ref{b1}), 
we see that $\alpha$ is determined by the detailed 
physics on micro scales of the bounce process. 
Finally, it should be emphasized that a specific case of 
the matter bounce scenario~\cite{B-R-BCS, Cai:2013kja} 
investigated by Brandenberger et al. 
is able to be realized also in $F(R)$ gravity.

%%%%%%%%%%%%%%%%%%%%%%%%%%%
%%%  Sec. IV
%%%%%%%%%%%%%%%%%%%%%%%%%%%
\section{Stability of the solutions}

In this section, with the procedure of the first reference in Ref.~\cite{R-F(R)-NO-CF-CD-DS}, 
we examine the stability of the solutions in 
$F(R)$ gravity models obtained in Sec.~III. 

We suppose that a solution of Eq.~(\ref{eq:2.2}) is 
expressed as $G= G_\mathrm{b} (N)$. 
The description of $G$ including 
the perturbation $\delta G (N)$ from the background solution $G_\mathrm{b} (N)$ is given by $G(N) = G_\mathrm{b} (N) + \delta G (N)$. 
%%%%%
Here, we note that $N (\geq 0)$ is equal to or larger than $0$. 
(This is clearly seen from the first equation in (\ref{b4}) with $\alpha >0$ 
and Eq.~(\ref{eq:3.10}) with $\bar{a} >0$.)
%%%%%
By substituting this 
expression into Eq.~(\ref{eq:2.2}), we find 
\begin{eqnarray}
\hspace{-5mm}
&&
J_1 \delta G^{''} (N) 
+ J_2 \delta G^{'} (N) + J_3 \delta G (N) =0\,, 
\label{eq:4.1} \\
\hspace{-5mm}
&&
J_1 \equiv G_\mathrm{b} (N) \frac{d^2 F(R)}{d R^2}\,, 
\label{eq:4.2} \\
\hspace{-5mm}
&&
J_2 \equiv 3G_\mathrm{b} (N) \left[ 
 \left(4G_\mathrm{b}^{'} (N) +G_\mathrm{b}^{''} (N) \right) \frac{d^3 F(R)}{d R^3} + \left(1- \frac{1}{6} \frac{G_\mathrm{b}^{'} (N)}{G_\mathrm{b} (N)}  \right) \frac{d^2 F(R)}{d R^2} \right]\,, 
\label{eq:4.3} \\
\hspace{-5mm}
&&
J_3 \equiv G_\mathrm{b} (N) \left[ 
12\left(4G_\mathrm{b}^{'} (N) +G_\mathrm{b}^{''} (N) \right) \frac{d^3 F(R)}{d R^3} 
%\right. \nonumber \\
%&& \left. 
%\hspace{25mm}
%{}
- \left(4 -2 \frac{G_\mathrm{b}^{'} (N)}{G_\mathrm{b} (N)} - \frac{G_\mathrm{b}^{''} (N)}{G_\mathrm{b} (N)} \right) \frac{d^2 F(R)}{d R^2} 
+ \frac{1}{3} \frac{1}{G_\mathrm{b} (N)} \frac{d F(R)}{d R} \right]\,,
\label{eq:4.4}
\end{eqnarray}
where the values of $d F(R)/d R$, $d^2 F(R)/d R^2$ 
and $d^3 F(R)/d R^3$ are the ones at 
$R=3G_\mathrm{b}^{'} (N) +12G_\mathrm{b} (N)$ following from 
Eq.~(\ref{b3}).
Thus, the stability conditions $J_2 / J_1 >0$ and $J_3 / J_1 >0$ 
can be written as 
\begin{eqnarray}
%\hspace{-10mm}
&&
6\left(4G_\mathrm{b}^{'} (N) +G_\mathrm{b}^{''} (N) \right) \frac{d^3 F(R)}{d R^3} \left(\frac{d^2 F(R)}{d R^2}\right)^{-1} 
+ \left(6-\frac{G_\mathrm{b}^{'} (N)}{G_\mathrm{b} (N)}  \right) >0\,, 
\label{eq:4.5} \\ 
%\hspace{-10mm}
&&
36\left(4G_\mathrm{b}^{'} (N) +G_\mathrm{b}^{''} (N) \right) \frac{d^3 F(R)}{d R^3} \left(\frac{d^2 F(R)}{d R^2}\right)^{-1} 
%\right. 
\nonumber \\
&& 
%\left. 
\hspace{0mm}
{}
- 3\left(4 -2 \frac{G_\mathrm{b}^{'} (N)}{G_\mathrm{b} (N)} - \frac{G_\mathrm{b}^{''} (N)}{G_\mathrm{b} (N)} \right) 
+ \frac{1}{G_\mathrm{b} (N)} \frac{d F(R)}{d R} 
\left(\frac{d^2 F(R)}{d R^2}\right)^{-1} >0\,.
\label{eq:4.6}
\end{eqnarray}
%

%%%%%%%%%%%%%%%%%%%%%%%%%%%
%%%  Sec. IV A
%%%%%%%%%%%%%%%%%%%%%%%%%%%
\subsection{Stability of the exponential model}

In the case that the scale factor is described by 
an exponential form in the exponential model, 
with Eqs.~(\ref{b3}), (\ref{b5}) and (\ref{b8}) we see that 
$G_\mathrm{b} = 4\alpha N$ and therefore 
the first condition in (\ref{eq:4.5}) reads 
$6-1/N >0$. Moreover, regarding 
the second condition in (\ref{eq:4.6}), the quantity on the 
left-hand side is equal to zero. In other words, 
the quantity $J_3 / J_1$ is not negative. 
Consequently, if $N<0$ or $N>1/6$, the solution could be stable. 
The latter condition can be satisfied because $N$ has to be 
much larger than unity. Thus, the exponential model of 
the scale factor could be stable.

%%%%%%%%%%%%%%%%%%%%%%%%%%%
%%%  Sec. IV B
%%%%%%%%%%%%%%%%%%%%%%%%%%%
\subsection{Stability of the power-law model}

When the scale factor has a power-law form, given by Eq.~(\ref{eq:3.9}),  
the first stability condition (\ref{eq:4.5}) becomes 
\begin{eqnarray}
&&
\frac{2}{\e^{N} -1} \left[ 
\frac{q}{2q-1} \left(\beta -2 \right) 
\left( 6-\frac{1}{q} -\frac{4\e^{N}}{q} +\frac{2\e^{N}}{q^2} -4 \e^{-N} 
\right) +3 \left(\e^{N} -1 \right) 
-\left(2-\frac{\e^{N}}{q} \right) 
\right] 
\nonumber \\
&&
\simeq 
\frac{2}{\e^{N} -1} \left[ \beta \left(1-\frac{2}{q} \right) 
+ \frac{5}{q} -4 \right] > 0\,,
\label{eq:3.19}  
\end{eqnarray}
whereas the second stability condition (\ref{eq:4.5}) reads
\begin{eqnarray}
&&
\frac{6}{\left(\e^{N} -1\right)^2} \left\{ 
\left( -2-\frac{5}{q} +\frac{2\e^{N}}{q^2} +4\e^{-N} \right)\e^{N} 
%\right. 
%\nonumber \\
%&& 
%\left. 
%{}
+\left[ 2\left(1-\frac{2}{q}\right)\left(\beta-2\right) 
+ 2\left(1-\frac{\e^{N}}{q}\right) + 
\frac{1}{\beta-1} \e^{N} \left(2 -\frac{1}{q}\right)
\right] \left(\e^{N} -1\right)
\right\}
\nonumber \\
&&
\simeq  
\frac{6}{\left(\e^{N} -1\right)^2} 
\frac{1}{q^2} \left(2q-1\right)\left(q-2\right)
> 0\,.
\label{eq:FR7-4-IVC-a1}
\end{eqnarray}
Here, in deriving Eqs.~(\ref{eq:3.19}) and (\ref{eq:FR7-4-IVC-a1}), 
we have used $\e^{N} \simeq 1$ in those numerators. {}From Eq.~(\ref{eq:3.19}), we see that if $\beta \left(1-2/q \right) + 5/q -4 >0$, 
the first stability condition can be satisfied. 
Furthermore, it follows from Eq.~(\ref{eq:FR7-4-IVC-a1}) 
that for $q<1/2$ or $q>2$, 
the second stability condition can be met.

%%%%%%%%%%%%%%%%%%%%%%%%%%%
%%%  Sec. V
%%%%%%%%%%%%%%%%%%%%%%%%%%%
\section{Unified $F(R)$ model of bounce and the late-time 
cosmic accelerated expansion}

In this section, we reconstruct an $F(R)$ model where not only 
the bouncing behavior in the early universe but also 
the late-time accelerated expansion of the universe 
can be realized in a unified manner.

%%%%%%%%%%%%%%%%%%%%%%%%%%%
%%%  Sec. V A 
%%%%%%%%%%%%%%%%%%%%%%%%%%%
\subsection{Sum of exponentials model} 

As a concrete model, 
we investigate a sum of exponentials form for the scale factor 
\begin{eqnarray} 
a(t) \Eqn{=} \e^{Y} + \e^{Y^2}\,, 
\label{eq:FR7-6-VA-01-1} \\ 
Y \Eqn{\equiv} \left(\frac{t}{\bar{t}}\right)^2\,.
\label{eq:FR7-6-VA-01-2}
\end{eqnarray}
Here, we again note that $\bar{t}$ is a fiducial time. 
In this model, for the limit $t/\bar{t} \to 0$, 
i.e., in the early universe, we obtain 
$a \to \e^{Y}$, which is equivalent to $a = \e^{\alpha t^2}$ with $\alpha = 1/\bar{t}^2$ in Eq.~(\ref{b1}), 
and hence the bouncing behavior can occur. 
While, in the limit $t/\bar{t} \gg 1$, we find $a \to \e^{Y^2}$ and 
hence $\ddot{a} = 4 \left(1/\bar{t}\right)^2 Y \left(3 + 4 Y^2 \right) \e^{Y^2} 
>0$. Consequently, the late-time accelerated expansion of the universe 
can be realized. 
In the following, we analyze cosmological quantities around $t =0$ 
in order to examine bounce cosmology. 
With $N = \ln a$ and $H = \dot{N}$, 
the form of $a$ in Eq.~(\ref{eq:FR7-6-VA-01-1}) leads to 
\begin{eqnarray} 
N \Eqn{=} 
\ln \left( \e^{Y} + \e^{Y^2} \right)
\approx \ln \left( 2 + Y + \frac{3}{2} Y^2 \right)\,,
\label{eq:FR7-6-VA-02}\\
H \Eqn{=} \frac{2\left( 1+3Y \right) \dot{Y}}{3Y^2 + 2Y +4} 
\approx \frac{\dot{Y}}{2}\,, 
\label{eq:FR7-6-VA-03}
\end{eqnarray}
where in deriving the approximate equalities in 
Eqs.~(\ref{eq:FR7-6-VA-02}) and (\ref{eq:FR7-6-VA-03}) 
we have expanded the exponential function in terms of $Y$ 
and used $Y \ll 1$. 
By solving the approximate equality in Eq.~(\ref{eq:FR7-6-VA-02}) 
with respect to Y and taking into account the fact that 
$Y = \left( t/\bar{t} \right)^2 >0$ 
as in Eq.~(\ref{eq:FR7-6-VA-01-2}), 
we acquire 
\begin{eqnarray} 
t \Eqn{=} \pm \sqrt{\frac{\sqrt{D} - 1}{3\left( 1/\bar{t} \right)^2}}\,,
\label{eq:FR7-6-VA-04} \\ 
D \Eqn{\equiv} 6 \e^{N} -11\,. 
\label{eq:FR7-6-VA-05}
\end{eqnarray}
Here, $D > 1$ because $t$ should be a real number. 
Thus, from this inequality we have $\e^{N} >2$, i.e., $N > \ln2$. 
This constraint on $N$ can be satisfied because $N \gg 1$. {}From $G=H^2$ and 
Eq.~(\ref{b3}), we find 
\begin{eqnarray} 
G(N) \Eqn{=} \frac{1}{3\bar{t}^2} \left( -1 + \sqrt{6 \e^{N} -11} \right)\,,
\label{eq:FR7-6-VA-06}\\
R \Eqn{=} \frac{2}{\bar{t}^2} \left( 1+2\sqrt{6 \e^{N} -11} \right)\,.
\label{eq:FR7-6-VA-07}
\end{eqnarray}
Accordingly, 
by applying Eqs.~(\ref{eq:FR7-6-VA-06}) and (\ref{eq:FR7-6-VA-07}) to Eq.~(\ref{b2}) and providing that contributions from matter are negligible, we acquire
\begin{equation}
-\frac{24}{\bar{t}^2} \left( R-\frac{6}{\bar{t}^2} \right) 
\frac{d^2 F(R)}{dR^2} 
+\left( R + \frac{6}{\bar{t}^2} \right) \frac{dF(R)}{dR} - 2F(R) = 0\,.
\label{eq:FR7-6-VA-08} 
\end{equation}
We find a solution of this equation as 
\begin{equation}
F(R) = \bar{t}^2 R^2 -36R +\frac{36}{\bar{t}^2}\,.
\label{eq:FR7-6-VA-09} 
\end{equation}
Here, the reason why the solution in Eq.~(\ref{eq:FR7-6-VA-09}) includes 
$\bar{t}$ is that the dimension of the $F(R)$ form is adjusted to be 
mass squared ($[\mathrm{Mass}]^2$). 

%%%
{}From Eq.~(\ref{eq:FR7-6-VA-09}) with Eqs.~(\ref{eq:FR7-6-VA-04}) and (\ref{eq:FR7-6-VA-07}), we acquire 
$F'(R) = 2\bar{t}^2 \left(R-18/\bar{t}^2\right) 
= 24 \left[\left(t/\bar{t}\right)^2 -1\right]$. 
Accordingly, when a bounce happens at $t =0$, $F'(R) <0$. 
We also see that $F'(R) = 0$ at $R = 18/\bar{t}^2$. 
Consequently, for the scale factor in Eq.~(\ref{eq:FR7-6-VA-01-1}), 
the bounce is realized in the regime when a effective 
gravitational constant is negative, namely, graviton is a ghost. 
On the other hand, since $F''(R) =2\bar{t}^2 > 0$, the cosmological 
perturbations can be stable~\cite{Nojiri:2003ft, DK-F-SHS} . 
%%%

\begin{center}
\begin{figure}[tbp]
\resizebox{!}{6.5cm}{
   \includegraphics{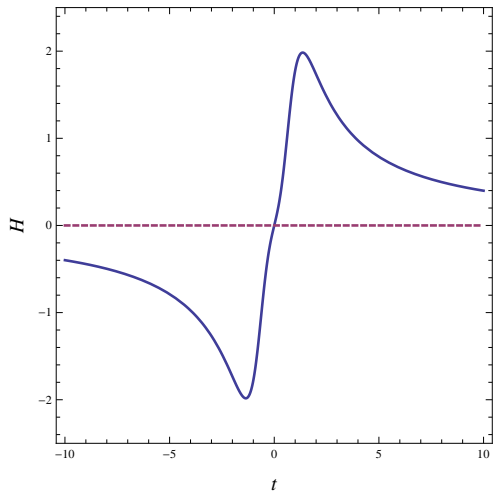}
                  }
\caption{The Hubble parameter in the first equality 
with $\bar{t}=1$ in Eq.~(\ref{eq:FR7-6-VA-03}) around a bounce at $t=0$. 
Legend is the same as Fig.~\ref{FR7-fig-2}. 
} 
\label{FR7-fig-3}
\end{figure}
\end{center}

%%%%%%%%
In Fig.~\ref{FR7-fig-3}, 
we display the behavior of the Hubble parameter in the first equality 
with $\bar{t}=1$ in Eq.~(\ref{eq:FR7-6-VA-03}) around a bounce at $t=0$. 
In this figure, before the bounce ($t<0$), we have $H <0$, 
and after it ($t>0$), we obtain $H>0$. 
This is the same behavior as Figs.~\ref{FR7-fig-1} and \ref{FR7-fig-2}, 
and therefore the bouncing behavior emerges. 
%%%%%%%%

%%%%%%%%%%%%%
%%% Fig-3 %%%
%%%%%%%%%%%%%%%%%%%%%%%%%%%%%%%%%%%%%%%%%%%%%%%%%%%%%%
%%%%%%%%%%%%%%%%%%%%%%%%%%%%%%%%%%%%%%%%%%%%%%%%%%%%%%

%%%%%%%%%%%%%%%%%%%%%%%%%%%
%%%  Sec. V B
%%%%%%%%%%%%%%%%%%%%%%%%%%%
\subsection{Stability of the sum of exponentials model}

For the double exponential model in Eq.~(\ref{eq:FR7-6-VA-01-1}), the stability condition (\ref{eq:4.5}) reads 
\begin{equation}
6 \left( \frac{-2 +\sqrt{6 \e^{N} -11}}{-1 +\sqrt{6 \e^{N} -11}} \right) > 0\,. \label{eq:FR7-6-VB-01}
\end{equation}
This is satisfied if $N >\ln\left(5/2\right)$. 
Since $N$ has to be much larger than unity, this condition can be met. 
Moreover, for Eq.~(\ref{eq:FR7-6-VA-01-1}), 
the left-hand side of the inequality (\ref{eq:4.6}) becomes zero. 
Presumably, if we include higher order term in $Y$, 
the left-hand side of the inequality (\ref{eq:4.6}) 
might be non-zero, and therefore that we can have some conditions 
on $N$, although it might be quite difficult to execute the nvestigations 
analytically. 
Hence, it would be expected that such a condition could be satisfied 
because of the large value of $N$. 
It follows from the above considerations that the sum of exponentials 
model could be compatible with the stability conditions.

%%%%%%%%%%%%%%%%%%%%%%%%%%%
%%%  Sec. VI 
%%%%%%%%%%%%%%%%%%%%%%%%%%% 
\section{Exponential form of the scale factor 
for the non-zero spatial curvature}

In Secs.~III A and IV A, we have seen that 
in the flat FLRW universe, 
an exponential form of the scale factor and 
the resultant second order polynomial model of $F(R)$ gravity 
could be a stable theory realizing the bounce cosmology. 
In this section, we examine an exponential form of the scale factor 
for the non-zero spatial curvature, namely, 
in the non-flat FLRW universe. 

A more general form of the FLRW metric is written as 
\begin{equation}
ds^2=-dt^2+a^2(t)\left(\frac{dr^2}{1-kr^2}+r^2d\theta^2+r^2\sin^2\theta d\phi^2\right)\,,
\label{eq:FR7-9-VI-01}
\end{equation}
where $k=0$ (flat universe), $+ 1$ (closed universe) and $-1$ (open universe) 
is the spatial curvature. 
The metric in Eq.~(\ref{eq:FR7-9-VI-01}) with $k=0$ is equivalent to 
that in Eq.~(\ref{eq:FR7-3-Add-01}). 
The action describing $F(R)$ gravity is given by (\ref{eq:2.1}) and in this case, the gravitational field equations read 
\begin{eqnarray}
\left(\frac{\Dot{a}}{a}\right)^2+\frac{k}{a^2} \Eqn{=} \frac{\kappa^2}{3F^{\prime}(R)}(\rho_\mathrm{M}+\rho_\mathrm{DE})\,,
\label{eqf1} \\ 
\frac{\Ddot{a}}{a} \Eqn{=} -\frac{\kappa^2}{6F^{\prime}(R)}(\rho_\mathrm{M}+3P_\mathrm{M}+\rho_\mathrm{DE}
+3P_\mathrm{DE})\,,
\label{eqf2}
\end{eqnarray}
where $\rho_\mathrm{DE}$ and $P_\mathrm{DE}$ are 
the energy density and pressure of dark energy components 
of the universe, respectively, defined by
\bea
\rho_\mathrm{DE}
\Eqn{\equiv}-\frac{1}{\kappa^2}\left(\frac{1}{2}F(R)-\frac{1}{2}RF^{\prime}(R)+3\frac{\Dot{a}}{a}\Dot{R}F^{\prime\prime}(R)\right)\,,\\
P_\mathrm{DE}
\Eqn{\equiv} 
\frac{1}{\kappa^2}\left[\frac{1}{2}F(R)-\frac{1}{2}RF^{\prime}(R)+
\left(2\frac{\Dot{a}}{a}\Dot{R}+\Ddot{R}\right)F^{\prime\prime}(R)+\Dot{R}^2F^{\prime\prime\prime}(R)\right]\,.
\eea
Here, the prime denotes the derivative with respect to the scalar curvature $R$ of $\partial/\partial R$. 

We examine the case that the scale factor is expressed as 
a linear combination of 
$\e^{\lambda t}$ and $\e^{-\lambda t}$, i.e., 
\begin{equation}
a(t)=\sigma \e^{\lambda t}+\tau \e^{-\lambda t}\,, 
\label{sol}
\end{equation}
with $\lambda$, $\sigma$ and $\tau$ constant real numbers  
($\lambda,\sigma,\tau\in\mathbb{R}$), 
$\tau \sigma \ne 0$ and $\lambda \ne 0$. 
%%%%%
We note that 
for $\tau = 0$ in Eq.~(\ref{sol}), $a \propto \e^{\lambda t}$, 
and hence such a metric describes the de Sitter solution 
with the Hubble parameter $H = \lambda$ when $k=0$. 
%%%%%
%%%%%%%%
Also, we mention that 
if $\sigma = \tau = 1/\left(2\lambda\right)$ for $k=+1$ or 
$\sigma = -\tau = 1/\left(2\lambda\right)$ for $k=+1$, 
we can have the de Sitter solution~\cite{H-E:1973}. 
%%%%%%%%
For this model, the corresponding expressions for the Hubble parameter and scalar curvature become 
\begin{eqnarray}
H(t) \Eqn{=} \frac{\Dot{a}}{a}=
\lambda\frac{\sigma \e^{\lambda t}-\tau \e^{-\lambda t}}{\sigma \e^{\lambda t}+\tau \e^{-\lambda t}}\,,
\label{hab} \\
R(t) \Eqn{=} \frac{6(a\Ddot{a}+\Dot{a}^2+k)}{a^2}=
\frac{6\left[2\lambda^2\left(\sigma^2 \e^{4\lambda t}+\tau^2\right)+k \e^{2\lambda t}\right]}{(\sigma \e^{2\lambda t}+\tau)^2}\,.
\label{rr}
\end{eqnarray}

%%%%%%%%%%%%%%%%%%%%%%%%%%%
%%%  Sec. VI A
%%%%%%%%%%%%%%%%%%%%%%%%%%% 
\subsection{Second order polynomial model}

As a form of $F(R)$ to realize the 
exponential model of the scale factor in Eq.~(\ref{sol}), 
we take a second order polynomial in terms of $R$ as 
\begin{equation}
F(R)=\alpha_0+\alpha_1R+\alpha_2R^2 \,, 
\label{fr1}
\end{equation}
where $\alpha_j$ with $j = 0, 1, 2$ are constant real numbers 
($\alpha_j \in\mathbb{R}$). 
By substituting Eqs.~(\ref{sol}), (\ref{rr}) and (\ref{fr1}) into 
the gravitational field equations (\ref{eqf1}) and (\ref{eqf2}), 
we find 
%
%\begin{multline}
\begin{eqnarray}
&&
(\alpha_0+6\alpha_1\lambda^2)\tau^4
+2\e^{2\lambda t}\tau^2\left[3k(\alpha_1-12\alpha_2\lambda^2)+2(\alpha_0+72\alpha_2\lambda^4)\sigma\tau\right] 
\nonumber \\
&& \hspace{10mm} 
{}+6\e^{4\lambda t}\left\{6k^2\alpha_2+2k(\alpha_1+12\alpha_2\lambda^2)\sigma\tau+[\alpha_0-2\lambda^2(\alpha_1+96\alpha_2\lambda^2)]\sigma^2\tau^2\right\}
\nonumber \\
&& \hspace{20mm}
{}+2\e^{6\lambda t}\sigma^2\left[3k(\alpha_1-12\alpha_2\lambda^2)+2(\alpha_0+72\alpha_2\lambda^4)\sigma\tau\right]+
\e^{8\lambda t}(\alpha_0+6\alpha_1\lambda^2)\sigma^4=0\,,
\label{eqfn1} \\ 
&&
(\alpha_0+6\alpha_1\lambda^2)\tau^4
+4\e^{2\lambda t}\sigma\tau^3(\alpha_0+6\alpha_1\lambda^2) 
\nonumber \\
&& \hspace{10mm} 
{}-6\e^{4\lambda t}\left\{6k^2\alpha_2+48k\alpha_2\lambda^2\sigma\tau-[\alpha_0+6\lambda^2(\alpha_1+48\alpha_2\lambda^2)]\sigma^2\tau^2\right\} 
\nonumber \\
&& \hspace{20mm} 
{}+4\e^{6\lambda t}(\alpha_0+6\alpha_1\lambda^2)\sigma^3\tau+\e^{8\lambda t}(\alpha_0+6\alpha_1\lambda^2)\sigma^4=0\,.
\label{eqfn2}
\end{eqnarray}
%\end{multline}
%
In addition, the following condition has to be satisfied 
\begin{equation}
(\alpha_1+24\alpha_2\lambda^2)\tau^2
+2\e^{2\lambda t}(6k\alpha_2+\alpha_1\sigma\tau)
+\e^{4\lambda t}(\alpha_1+24\alpha_2\lambda^2)\sigma^2\neq 0.
\end{equation}
It follows from Eqs.~(\ref{eqfn1}) and (\ref{eqfn2}) that 
we find the conditions in terms of the coefficients 
\begin{eqnarray}
&&
\alpha_0+6\alpha_1\lambda^2=0\,, \nonumber \\ 
&&
3k(\alpha_1-12\alpha_2\lambda^2)+2(\alpha_0+72\alpha_2\lambda^4)\sigma\tau=0\,, \nonumber \\ 
&& 
6k^2\alpha_2+2k(\alpha_1+12\alpha_2\lambda^2)\sigma\tau +
[\alpha_0-2\lambda^2(\alpha_1+96\alpha_2\lambda^2)]\sigma^2\tau^2=0\,, 
\nonumber \\
&& 
6k^2\alpha_2+48k\alpha_2\lambda^2\sigma\tau-\left[\alpha_0+6\lambda^2(\alpha_1+48\alpha_2\lambda^2)\right]\sigma^2\tau^2=0\,. \nonumber
\end{eqnarray}
%\end{gather*}
%
These equations are rewritten to 
%\begin{gather}
\begin{eqnarray}
&& 
\alpha_0+6\alpha_1\lambda^2=0\,, \nonumber \\
&& 
(\alpha_1-12\alpha_2\lambda^2)(k-4\lambda^2\sigma\tau)=0\,, \nonumber \\
&& 
\left[3k\alpha_2+(\alpha_1+24\alpha_2\lambda^2)\sigma\tau\right](k-4\lambda^2\sigma\tau)=0\,, 
\label{eqfnn} \\ 
&&
\alpha_2(k+12\lambda^2\sigma\tau)(k-4\lambda^2\sigma\tau)=0\,. \nonumber 
\end{eqnarray}
%\end{gather}
%
For $\alpha_0 \alpha_1 \alpha_2\ne0$, 
from the system of equations in (\ref{eqfnn}) we have two different sets of 
the conditions on the parameters: 
(a) 
%$\noindent 1.\ \ \ 
$\alpha_0+6\alpha_1\lambda^2=0$, \, 
$k-4\lambda^2\sigma\tau=0$, 
and 
(b) 
%\noindent 2.\ \ \ 
$\alpha_0+6\alpha_1\lambda^2=0$, \, 
$\alpha_1-12\alpha_2\lambda^2=0$, \, 
$k+12\lambda^2\sigma\tau=0$. 
In both cases, we acquire 
$\lambda=\pm\sqrt{-\frac{\alpha_0}{6\alpha_1}}$. 
Without loss of generality, we can assume that $\lambda>0$ and $\sigma>0$. 
In this case, the set of solutions of the gravitational field equations in the 
FLRW universe is divided into the following three types. 

\begin{itemize}
\item 
{\it Type I} 
%\noindent{\it Type I} 
%$$
\begin{equation}
\alpha_0\alpha_1<0\,, \quad \alpha_2\neq\frac{\alpha_1^2}{4\alpha_0}\,, \quad \lambda=\sqrt{-\frac{\alpha_0}{6\alpha_1}}\,, 
\quad \sigma>0\,, \quad \tau=\frac{k}{4\lambda^2\sigma}\,, \quad k=\pm 1\,. 
\nonumber 
\end{equation}
%$$
{}From this set of parameters, we see that
\begin{equation}
R=-\frac{2\alpha_0}{\alpha_1}\,, \quad w_\mathrm{DE}=-1\,,
\nonumber 
\end{equation}
%$$R=-\frac{2\alpha_0}{\alpha_1},\quad w_{eff}=-1.$$
where $w_\mathrm{DE}$ is the equation of state of the dark energy component 
defined by $w_\mathrm{DE} \equiv P_\mathrm{DE}/\rho_\mathrm{DE}$.

\item 
{\it Type II} 
%\noindent{\it Type II}
%$$
\begin{equation}
\alpha_0\alpha_1<0\,,\quad \alpha_2=-\frac{\alpha_1^2}{2\alpha_0}\,,\quad \lambda=\sqrt{-\frac{\alpha_0}{6\alpha_1}}\,,
\quad \sigma>0\,,\quad \tau=\frac{-k}{12\lambda^2\sigma}\,,\quad k=\pm 1\,.
\nonumber 
\end{equation}
%$$
{}From this set of parameters, we find that
%$$
\begin{equation}
R=-\frac{2\alpha_0}{\alpha_1}\left[1+\frac{96 \e^{2\lambda t}k\lambda^2\sigma^2}{(k-12 \e^{2\lambda t}\lambda^2\sigma^2)^2}\right]\,,
\quad w_\mathrm{DE} =-1+f(k,\sigma,\tau,\lambda,\alpha_0,\alpha_1,\alpha_2)\,, 
\nonumber 
\end{equation}
%$$
where $f$ is a function of the parameters 
$k$, $\sigma$, $\tau$, $\lambda$, $\alpha_0$, $\alpha_1$, and $\alpha_2$. 

\item 
{\it Type III} 
%\noindent{\it Type III}
%$$
\begin{equation}
\alpha_0=0\,,\quad \alpha_1=0\,,\quad \alpha_2\neq 0\,,\quad \lambda>0,
\quad \sigma>0\,,\quad \tau=\frac{k}{4\lambda^2\sigma}\,,\quad k=\pm 1\,.
\nonumber 
\end{equation}
%$$
{}From this set of parameters, we obtain 
%$$
\begin{equation}
R=12\lambda^2\,, \quad w_\mathrm{DE}=-1\,.
\nonumber 
\end{equation}
%$$
\end{itemize}
It should be noted that for 
the form of the function $F(R)$ in Eq.~(\ref{fr1}), 
there is no solution other than the de Sitter solution, 
if the cosmic curvature $k$ is zero (and also $f=0$). 

We also remark that 
as the scale factor $a(t)$ satisfying the above solutions, 
a more general expression can be described by replacing 
$t\rightarrow t-t_1$ with $t_1$ another fiducial time, i.e., 
%$$
\begin{equation}
a(t)=\sigma \e^{\lambda(t-t_1)}+\tau \e^{-\lambda(t-t_1)}\,.
\nonumber 
\end{equation}
%$$
Similarly, $F(R)$ can be generalized as any function of the form 
\be
F(R)=\beta_l\frac{1}{R^l}+...+\beta_1\frac{1}{R}+\alpha_0+\alpha_1R+...+\alpha_mR^m
\,,
\label{pol1}
\ee
where $\beta_j$ ($j = 1, \dots, l$) and $\alpha_i$ ($i = 0, \dots, m$) are
constants.

%%%%%%%%%%%%%%%%%%%%%%%%%%%
%%%  Sec. VI B
%%%%%%%%%%%%%%%%%%%%%%%%%%% 
\subsection{Model consisting of an inverse power-law term}  

Next, we investigate the function $F(R)$ expressed as~\cite{Carroll:2003wy} 
\be 
F(R)=\alpha_1R+\beta_1\frac{1}{R}\,.
\label{eq:FR7-19-VIB-ADD-01}
\ee
With the similar procedure developed in the preceding subsection, 
we obtain the following restrictions on the parameters 
%$$
\begin{equation}
\beta_1+48\alpha_1\lambda^4=0\,,\quad k-4\lambda^2\sigma\tau=0\,.
\nonumber 
\end{equation}
%$$
In addition, the following condition has to be met 
%$$
\begin{equation}
(\beta_1-144\alpha_1\lambda^4)^2+(-36k\alpha_1\lambda^2+\beta_1\sigma\tau)^2+
\left[6k^2\alpha_1-(\beta_1-48\alpha_1\lambda^4)\sigma^2\tau^2\right]^2\neq 
0\,. 
\nonumber 
\end{equation}
%$$
It is easy to rewrite this equation in the following form 
%$$
\begin{equation}
9k^4\alpha_1^2-3k^2\alpha_1\beta_1+\beta_1^2\neq 0\,.
\nonumber 
\end{equation}
%$$
{}From this equation, we obtain the restrictions on the parameters 
%$$
\begin{equation}
\alpha_1\beta_1<0\,, \quad \lambda=\sqrt[4]{-\frac{\beta_1}{48\alpha_1}}>0\,, 
\quad \sigma>0\,, \quad 
\tau=\frac{k}{4\lambda^2\sigma}\,, \quad k=\pm 1\,.
\nonumber 
\end{equation}
%$$
{}From this set of parameters, we see that 
%$$
\begin{equation}
R=12\sqrt{-\frac{\beta_1}{48\alpha_1}}\,, \quad w_\mathrm{DE}=-1\,.
\nonumber 
\end{equation}

%%%%%%%%
It should be cautioned that 
in the model in Eq.~(\ref{eq:FR7-19-VIB-ADD-01}), 
the late-time cosmic acceleration which is accepted from the quantum field theoretical point of view cannot be realized because its de Sitter solution exists 
in the unstable region where $F''(R) < 0$.  
%%%%%%%%

%%%%%%%%
We can consider a slightly different form of the function $F(R)$ as 
\begin{equation}
F(R)=\alpha_0+\alpha_1R+\beta_1\frac{1}{R}\,.
\end{equation}
With the similar procedure developed in the preceding subsection, we obtain the following restrictions on the parameters 
\begin{equation}
\beta_1+8\alpha_0\lambda^2+48\alpha_1\lambda^4=0, \quad 
k-4\lambda^2\sigma\tau=0\,.
\end{equation}
In addition, the following condition has to be met 
\begin{equation}
(\beta_1-144\alpha_1\lambda^4)^2+(\beta_1\sigma\tau-36k\alpha_1\lambda^2)^2+
(6k^2\alpha_1-(\beta_1-48\alpha_1\lambda^4)\sigma^2\tau^2)^2\neq 0\,.
\end{equation}
It is easy to rewrite this equation in the following form 
\begin{equation}
16(\beta_1+6\alpha_0\lambda^2)^2+4k^2(6k\alpha_1+\alpha_0\sigma\tau)^2+(6k\alpha_0+4\beta_1\sigma\tau)^2\neq 0\,.
\end{equation}
For this type of a function $F(R)$, we have more complicated solutions, but we can impose additional restrictions and find a set of parameters for which 
$F''(R)>0$. 
For example, 
\begin{equation}
\alpha_1>0\,,\quad \beta_1>0\,,\quad \alpha_0=-2\sqrt{\alpha_1\beta_1}\,, 
\quad \lambda=\frac{1}{2}\left(\frac{\beta_1}{\alpha_1}\right)^{1/4}\,, 
\quad \sigma>0\,, \quad \tau=\frac{k}{4\lambda^2\sigma}\,, \quad k=\pm1\,, 
\end{equation}
or
\begin{equation}
\alpha_1>0\,, \quad \beta_1>0\,, 
\quad \alpha_0=-\sqrt{3}\sqrt{\alpha_1\beta_1}\,, 
\quad \lambda=\frac{1}{2}\left(\frac{\beta_1}{3\alpha_1}\right)^{1/4}\,, 
\quad \sigma>0\,, \quad \tau=\frac{k}{4\lambda^2\sigma}\,, \quad k=\pm1\,.
\end{equation}
{}From this set of parameters, we see that 
\begin{equation}
R=12\lambda^2,\quad  w_{\mathrm{DE}}=-1\,.
\end{equation}
%
%%%%%%%%

%%%%%%%%
In summary, in this section, for the FLRW universe with non-zero spatial curvature, when the scale factor is given by an exponential form in Eq.~(\ref{sol}), 
we have reconstructed 
a second order polynomial $F(R)$ model in terms of $R$ 
and an $F(R)$ model consisting of both a term proportional to $R$ 
and an inverse power-law term. 
It has been found that the de Sitter solution can exist for the case 
with non-zero spatial curvature. 
%%%%%%%%
Related to these consequences, 
as noted in Introduction, we again mention that 
if the spatial curvature is positive, i.e., $k (>0)$, 
and a massive scalar field exists, 
the scale factor as well as the Riemann curvature can perform 
the bouncing behaviors~\cite{P-S-BS}. 
Moreover, when the spatial curvature has a non-zero value, 
namely, $k (\neq 0)$, 
in the Starobinsky model~\cite{Starobinsky:1980te} 
there is a solution where the scale factor can behave a bounce. 
%%%%%%%%

%%%%%%%%%%%%%%%%%%%%%%%%%%%
%%%  Sec. VII 
%%%%%%%%%%%%%%%%%%%%%%%%%%%
\section{Exponential form of the scale factor for the zero spatial curvature}

In the study of the bouncing behavior with an exponential form 
of the scale factor, it seems that another version of the reconstruction 
method (with an auxiliary scalar field) is more suitable. 
Hence, in this section we apply it to the derivation of $F(R)$ gravity models 
realizing bounce cosmology.

%%%%%%%%%%%%%%%%%%%%%%%%%%%
%%%  Sec. VII A
%%%%%%%%%%%%%%%%%%%%%%%%%%%
\subsection{Reconstruction method of $F(R)$ gravity}

When the scale factor is given by an exponential form in Eq.~(\ref{sol}), 
with the reconstruction method~\cite{odin1}, 
we find $F(R)$ gravity models with realizing the bounce cosmology. 
By using proper functions $P(t)$ and $Q(t)$ of a scalar field $t$ which we identify with the cosmic time, the action in Eq.~(\ref{eq:2.1}) can be represented 
as 
\begin{equation}
S=\frac{1}{2 \kappa^2}\int\sqrt{-g}\left(P(t)R+Q(t)\right)d^4 x \,.
\label{act}
\end{equation}
The variation with respect to $t$ yields
\begin{equation}
\frac{dP(t)}{dt}R+\frac{dQ(t)}{dt}=0\,, 
\label{eqvc}
\end{equation}
from which it is possible to solve $t$ in terms of $R$ as $t=t(R)$. 
By substituting $t=t(R)$ into Eq.~(\ref{act}), 
$F(R)$ can be written as 
\begin{equation}
F(R)=P(t(R))R+Q(t(R))\,.
\label{prq}
\end{equation}
With Eq.~(\ref{eqf1}), $Q(t)$ is given by
\begin{equation}
Q(t)=-6H^2(t)P(t)-6H(t)\frac{dP(t)}{dt}\,.
\label{funcQ}
\end{equation} 
Taking into account Eq.~(\ref{funcQ}), from Eq.~(\ref{eqf2}) we have the differential equation 
\begin{equation}
\frac{d^2P(t)}{dt^2}-H(t)\frac{dP(t)}{dt}+2\Dot{H}(t)P(t)=0\,,
\label{funcP}
\end{equation}
where we have used the expression of the Hubble parameter 
$H=\dot{a}/a$ of the first equality in (\ref{hab}). 
There are two different cases.

%%%%%%%%%%%%%%%%%%%%%%%%%%%
%%%  Sec. VII A 1
%%%%%%%%%%%%%%%%%%%%%%%%%%%
\subsubsection{Case 1: $\lambda>0$, $\sigma>0$, $\tau>0$}

\begin{center}
\begin{figure}[tbp]
\resizebox{!}{6.5cm}{
   \includegraphics{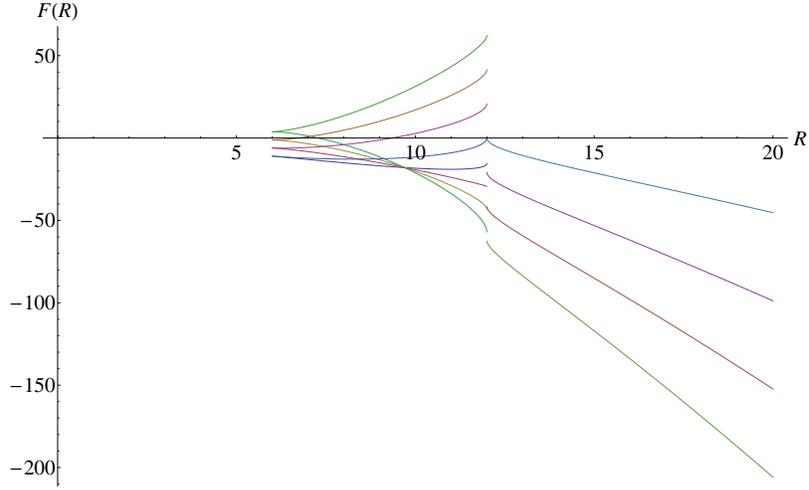}
                  }
\caption{
$F^{(1)}_{\pm}(R)$ ($6\lambda^2\leq R<12\lambda^2$) 
and $F^{(2)}_{+}(R)$ ($R>12\lambda^2$) 
as a function of $R$ with the parameters 
$c_1=1$, $c_2=0;\ 1 ;\ 2 ;\ 3$ (from the bottom to the top), 
$\lambda=1$, $\sigma=1$, $\tau=1$ for $F^{(1)}_{\pm}(R)$ 
and $\tau=-1$ for $F^{(2)}_{+}(R)$.
} 
\label{graph_1}
\end{figure}
\end{center}

The general solution of Eq.~(\ref{funcP}) is given by 
%$$
\begin{equation}
P(t)=(\sigma \e^{\lambda t}+\tau \e^{-\lambda t})\left[c_1\cos\left(2\sqrt{3}
\arctan\left(\e^{\lambda t}\sqrt{\frac{\sigma}{\tau}}\ \right)\right)+
c_2\sin\left(2\sqrt{3}\arctan\left(\e^{\lambda t}\sqrt{\frac{\sigma}{\tau}}\ \right)\right)\right]\,,
\nonumber
\end{equation}
%$$
where $c_1$ and $c_2$ are constants. {}From Eq.~(\ref{funcQ}), we have 
%\begin{multline*}
\begin{eqnarray}
Q(t) \Eqn{=} -12\lambda^2\frac{\e^{2\lambda t}\sigma-\tau}{\e^{2\lambda t}\sigma+\tau} 
\left\{\left[c_1(\sigma \e^{\lambda t}-\tau \e^{-\lambda t})+\sqrt{3}c_2\sqrt{\sigma\tau}\right]\cos\left(2\sqrt{3}\arctan\left(\e^{\lambda t}\sqrt{\frac{\sigma}{\tau}}\ \right)\right)\right. \nonumber \\
&& 
{}+\left.\left[c_2(\sigma \e^{\lambda t}-\tau \e^{-\lambda t})-\sqrt{3}c_1\sqrt{\sigma\tau}\right]\sin\left(2\sqrt{3}\arctan\left(\e^{\lambda t}\sqrt{\frac{\sigma}{\tau}}\ \right)\right)\right\}\,.
\end{eqnarray}
%\end{multline*}
It follows from Eq.~(\ref{eqvc}) that 
%$$
\begin{equation}
t_{\pm}=\frac{1}{2\lambda}\ln\left[\frac{-R\tau\pm 2\sqrt{6}\lambda\tau\sqrt{R-6\lambda^2}}{(R-12\lambda^2)\sigma}\right]\,, 
\quad 
6\lambda^2\leq R<12\lambda^2 \,.
\nonumber
\end{equation}
%$$
By solving Eq.~(\ref{prq}), we find the most general form of $F(R)$ 
\begin{equation*}
F^{(1)}_{\pm}(R)=2\sqrt{6}\lambda\sqrt{\sigma\tau}\left(A^{(1)}_{\pm}(R)\cos C^{(1)}_{\pm}(R)+
B^{(1)}_{\pm}(R)\sin C^{(1)}_{\pm}(R)\right)\,,
\end{equation*}
where 
%$$
\begin{eqnarray}
A^{(1)}_{\pm}(R) \Eqn{=} 
\pm\sqrt{3}c_2\sqrt{R-6\lambda^2}+c_1\sqrt{12\lambda^2-R}\,, 
\nonumber \\ 
%\quad 
B^{(1)}_{\pm}(R) \Eqn{=} \mp\sqrt{3}c_1\sqrt{R-6\lambda^2}+c_2\sqrt{12\lambda^2-R}\,, \nonumber \\ 
C^{(1)}_{\pm}(R) \Eqn{=} 2\sqrt{3}\arctan\left(\frac{\sqrt{6}\lambda\mp\sqrt{R-6\lambda^2}}{\sqrt{12\lambda^2-R}}\right)\,. 
\nonumber
\end{eqnarray}
%$$
Note that functions $F^{(1)}_{+}(R)$ and $F^{(1)}_{-}(R)$ are 
defined for the range $6\lambda^2\leq R<12\lambda^2$.
At the boundaries of the domain, 
these functions are characterized by the following behavior
%$$
\begin{eqnarray}
\lim_{R\rightarrow 6\lambda^2+0}F^{(1)}_{\pm}(R) \Eqn{=} 
12\lambda^2\sqrt{\sigma\tau}\left(c_1\cos\frac{\sqrt{3}}{2}\pi+
c_2\sin\frac{\sqrt{3}}{2}\pi\right)\,, \nonumber \\ 
\lim_{R\rightarrow 12\lambda^2-0}F^{(1)}_{+}(R) \Eqn{=} 
12\sqrt{3}\lambda^2\sqrt{\sigma\tau}c_2 \,, \nonumber \\ 
\lim_{R\rightarrow 12\lambda^2-0}F^{(1)}_{-}(R) \Eqn{=} 
12\sqrt{3}\lambda^2\sqrt{\sigma\tau}\left(c_1\sin\sqrt{3}\pi-
c_2\cos\sqrt{3}\pi\right)\,.
\nonumber
\end{eqnarray}
Also, we mention that the function $F^{(1)}_{-}(R)$ is fixed by the constants $c_1$, $c_2$, $\lambda$, $\sigma$, $\tau$. We acquire a central family of curves, the abscissa $R_0$ of the point intersection of curves of this family belongs to the region $6\lambda^2\leq R<12\lambda^2$ and determined from the equation 
\begin{equation}
\sqrt{12\lambda^2-R_0}
\sin C^{(1)}_{-}(R_0)-\sqrt{3}\sqrt{R_0-6\lambda^2}\cos C^{(1)}_{-}(R_0)=0\,. 
\nonumber
\end{equation}
A similar situation holds for the function $F^{(1)}_{+}(R)$. 
The equation for $R_0$ has the form 
\begin{equation}
A^{(1)}_{\pm}(R_0)\cos C^{(1)}_{\pm}(R_0)+B^{(1)}_{\pm}(R_0)\sin C^{(1)}_{\pm}(R_0)=0\,. 
\nonumber
\end{equation}
In Fig.~\ref{graph_1}, we depict 
$F^{(1)}_{\pm}(R)$ ($6\lambda^2\leq R<12\lambda^2$) as a function of $R$ 
with the parameters  
$c_1=1$, $c_2=0;\ 1 ;\ 2 ;\ 3$, 
$\lambda=1$, $\sigma=1$ and $\tau=1$.

%%%%%%%%%%%%%
%%%Fig.~4 %%%
%%%%%%%%%%%%%%%%%%%%%%%%%%%%%%%%%%%%%%%%%%%%%%%%%%%%%%
%%%%%%%%%%%%%%%%%%%%%%%%%%%%%%%%%%%%%%%%%%%%%%%%%%%%%%

%%%%%%%%%%%%%%%%%%%%%%%%%%%
%%%  Sec. VII A 2
%%%%%%%%%%%%%%%%%%%%%%%%%%%
\subsubsection{Case 2: $\lambda>0$, $\sigma>0$, $\tau<0$}

The general solution of Eq.~(\ref{funcP}) is given by
\begin{equation*}
P(t)=(\sigma \e^{\lambda t}+\tau \e^{-\lambda t})\left[c_1\cosh\left(2\sqrt{3}
\mathrm{arctanh}\left(\e^{\lambda t}\sqrt{-\frac{\sigma}{\tau}}\ 
\right)\right)+
c_2\sinh\left(2\sqrt{3}\mathrm{arctanh}\left(\e^{\lambda t}\sqrt{-\frac{\sigma}{\tau}}\ \right)\right)\right]\,. 
\end{equation*} 
{}From Eq.~(\ref{funcQ}), we obtain 
%\begin{multline*}
\begin{eqnarray}
Q(t) \Eqn{=} -12\lambda^2\frac{\e^{2\lambda t}\sigma-\tau}{\e^{2\lambda t}\sigma+\tau}
\left\{\left[c_1(\sigma \e^{\lambda t}-\tau \e^{-\lambda t})-\sqrt{3}c_2\sqrt{-\sigma\tau}\right]\cosh\left(2\sqrt{3}\mathrm{arctanh}\left(\e^{\lambda t}\sqrt{-\frac{\sigma}{\tau}}\ \right)\right)\right. \nonumber \\
&&
{}+\left.\left[c_2(\sigma \e^{\lambda t}-\tau \e^{-\lambda t})-\sqrt{3}c_1\sqrt{-\sigma\tau}\right]\sinh\left(2\sqrt{3}\mathrm{arctanh}\left(\e^{\lambda t}\sqrt{-\frac{\sigma}{\tau}}\ \right)\right)\right\}\,.
\nonumber
\end{eqnarray}
%\end{multline*}
{}From Eq.~(\ref{eqvc}), we obtain 
\begin{equation}
t_{\pm}=\frac{1}{2\lambda}\ln\left[\frac{-R\tau\pm 2\sqrt{6}\lambda\tau\sqrt{R-6\lambda^2}}{(R-12\lambda^2)\sigma}\right]\,, 
\quad 
R>12\lambda^2 \,.
\nonumber
\end{equation}
By solving Eq.~(\ref{prq}), we acquire the most general form of $F(R)$ 
\begin{equation*}
F^{(2)}_{\pm}(R)=2\sqrt{6}\lambda\sqrt{-\sigma\tau}\left(A^{(2)}_{\pm}(R)\cosh C^{(2)}_{\pm}(R)+
B^{(2)}_{\pm}(R)\sinh C^{(2)}_{\pm}(R)\right)\,,
\end{equation*}
where 
\begin{eqnarray}
A^{(2)}_{\pm}(R) \Eqn{=} \pm c_1\sqrt{R-12\lambda^2}\mp\sqrt{3}c_2\sqrt{R-6\lambda^2}\,,  \nonumber \\ 
B^{(2)}_{\pm}(R) \Eqn{=}\pm
c_2\sqrt{R-12\lambda^2}\mp\sqrt{3}c_1\sqrt{R-6\lambda^2}\,, \nonumber \\
C^{(2)}_{\pm}(R) \Eqn{=} 2\sqrt{3}\mathrm{arctanh}\left[\frac{\mp\sqrt{6}\lambda+\sqrt{R-6\lambda^2}}{\sqrt{R-12\lambda^2}}\right]\,.
\nonumber
\end{eqnarray}
We caution that $F^{(2)}_{-}(R)$ has no real values for $R>12\lambda^2$. 

At the boundaries of the domain, 
this function is characterized by the following behavior 
\begin{equation}
\lim_{R\rightarrow 12\lambda^2-0}F^{(2)}_{+}(R)=-12\sqrt{3}\lambda^2\sqrt{-\sigma\tau}c_2 \,.
\nonumber
\end{equation}
Hence, we have executed a reconstruction of $F(R)$ gravity for the scale factor in Eq.~(\ref{sol}), so that we have been able to build several types of 
$F(R)$ gravity theories realizing bounce cosmology. 
In Fig.~\ref{graph_1}, we plot 
$F^{(2)}_{+}(R)$ ($R>12\lambda^2$) as a function of $R$ 
with the parameters  
$c_1=1$, $c_2=0;\ 1 ;\ 2 ;\ 3$, 
$\lambda=1$, $\sigma=1$, and $\tau=-1$.

%%%%%%%%%%%%%%%%%%%%%%%%%%%
%%%  Sec. VII B 
%%%%%%%%%%%%%%%%%%%%%%%%%%%
\subsection{Stability of the solutions}

Next, we explore the stability of the obtained models. 
However, there are several problems associated with the large arbitrariness in the choice of the coefficients and the unwieldy of emerging relations. As an example, we study the stability of solutions $F_{-}^{(1)}(R)$ for one specific form of the metric. 

For instance, we consider a bouncing solution in the form
\begin{equation}
a(t)=\frac{1}{2}\e^{\lambda t}+\frac{1}{2}\e^{-\lambda t}=\cosh(\lambda t).
\label{stab1}
\end{equation}
For this model, we find  
\begin{equation}
N=\ln\cosh(t),\quad H=\Dot{N}=\lambda\tanh(\lambda t)\,, 
\label{eq:FR7-19-VII-B-ADD-01}
%\nonumber 
\end{equation}
which presents 
\begin{equation}
G(N)=H^2(N)=\lambda^2\left(1-\e^{-2N}\right)\,, \quad 
R=3G^{\prime}(N)+12G(N)=6\lambda^2\left(2-\e^{-2N}\right)\,. 
\nonumber 
\end{equation}
For the scale factor in Eq.~(\ref{stab1}), 
the stability conditions (\ref{eq:4.5}) and (\ref{eq:4.6}) can be written as 
follows. 

\begin{itemize}
\item  
%\noindent 
Case I ($c_1=0$, $c_2 \neq 0$) 

\begin{eqnarray}
&&
6-\frac{2}{-1+\e^{2 N}}+
\frac{2\sqrt{3}\e^{-N}}{\left(-1+\e^{2N}\right)
\left(1+\sqrt{1-\e^{-2N}}\right)} 
\frac{A\cos C+B\sin C}{D\cos C+E\sin C}>0\,, 
\nonumber \\ 
&&
-\frac{12\e^{-N}}{\left(-1+\e^{2N}\right)\left(1+\sqrt{1-\e^{-2N}}\right)}
\frac{\Bar{A}\cos C+\Bar{B}\sin C}{D\cos C+E\sin C}>0\,,
\nonumber 
\end{eqnarray}

\item  
%\noindent 
Case II ($c_1 \neq 0$, $c_2=0$)

\begin{eqnarray}
&&
6-\frac{2}{-1+\e^{2 N}}+
\frac{2\sqrt{3}\e^{-N}}{\left(-1+\e^{2N}\right)
\left(1+\sqrt{1-\e^{-2N}}\right)}
\frac{B\cos C-A\sin C}{-E\cos C+D\sin C}>0\,, \nonumber \\
&&
\frac{12\e^{-N}}{\left(-1+\e^{2N}\right)\left(1+\sqrt{1-\e^{-2N}}\right)}
\frac{\Bar{B}\cos C-\Bar{A}\sin C}{-E\cos C+D\sin C}>0\,,
\nonumber 
\end{eqnarray}
where 
\begin{eqnarray}
A \Eqn{=} \left(-4+3\e^{2N}\right)\left(-3\e^N+4\e^{3N}-\sqrt{-1+\e^{2N}}+4\e^{2N}\sqrt{-1+\e^{2N}}\right)\,, \nonumber \\ 
\Bar{A} \Eqn{=} 
-\sqrt{3}\left[\left(4-19\e^{2N}+12\e^{4N}\right)\sqrt{-1+\e^{2N}}-\e^N\left(-9+\sqrt{1-\e^{-2N}}\right) \right. \nonumber \\
&& 
\left. 
{}-4\e^{5N}\left(-2+\sqrt{1-\e^{-2N}}\right)+\e^{3N}\left(-18+5\sqrt{1-\e^{-2N}}\right)\right]\,, \nonumber \\ 
B \Eqn{=} \sqrt{3}\left(-2+\e^{2N}\right)\left[1+4\e^{4N}\left(1+\sqrt{1-\e^{-2N}}\right)-\e^{2N}\left(5+3\sqrt{1-\e^{-2N}}\right)\right]\,, \nonumber \\ 
\Bar{B} \Eqn{=}
5+\left(\e^N\sqrt{-1+e^{2N}}-3\e^{3N}\sqrt{-1+\e^{2N}}+2\e^{5N}
\sqrt{-1+\e^{2N}}
-10\e^{6N}\right)\left(1+\sqrt{1-\e^{-2N}}\right) \nonumber \\ 
&&
{}-13\e^{2N}\left(2+\sqrt{1-\e^{-2N}}\right)+\e^{4N}\left(31+25\sqrt{1-\e^{-2N}}\right)\,, \nonumber \\ 
D \Eqn{=} \sqrt{3}\left[-1+2\e^{2N}\left(1+\sqrt{1-\e^{-2N}}\right)\right]\,, 
\nonumber \\ 
E \Eqn{=} -2\e^N+2\e^{3N}-\sqrt{-1+\e^{2N}}+2\e^{2N}\sqrt{-1+\e^{2N}}\,,
\nonumber \\ 
C \Eqn{=} 2\sqrt{3}\arctan\left(\e^N+\sqrt{-1+\e^{2N}}\right)\,. 
\nonumber 
\end{eqnarray}

\end{itemize}

As a result, for case I, 
if $N>0.251224$, whereas for case II, 
when $N>0.0701889$, 
both stability conditions can be met. 
Since the value of $N$ is much larger than unity, 
these stability conditions can be satisfied. 
Thus, we find that for the scale factor in Eq.~(\ref{sol}), 
the model $F_{-}^{(1)}(R)$ is stable. 

\begin{center}
\begin{figure}[tbp]
\resizebox{!}{6.5cm}{
   \includegraphics{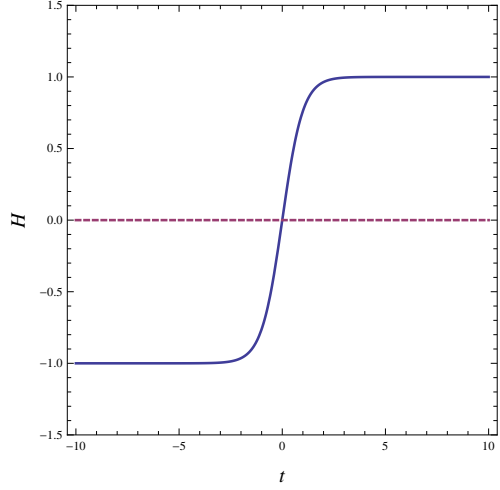}
                  }
\caption{The Hubble parameter in the second relation with 
$\lambda=1$ in (\ref{eq:FR7-19-VII-B-ADD-01}) around a bounce at $t=0$. 
Legend is the same as Fig.~\ref{FR7-fig-2}. 
} 
\label{FR7-fig-5}
\end{figure}
\end{center}

%%%%%%%%
In Fig.~\ref{FR7-fig-5}, 
we illustrate the behavior of the Hubble parameter in the second relation with 
$\lambda=1$ in (\ref{eq:FR7-19-VII-B-ADD-01}) around a bounce at $t=0$. {}From 
this figure, it is observed that before the bounce ($t<0$), $H <0$, 
and after it ($t>0$), $H>0$. 
This behavior is the same as Figs.~\ref{FR7-fig-1}--\ref{FR7-fig-3}. 
Accordingly, the bouncing behavior is realized. 
%%%%%%%%

%%%%%%%%%%%%%
%%% Fig-5 %%%
%%%%%%%%%%%%%%%%%%%%%%%%%%%%%%%%%%%%%%%%%%%%%%%%%%%%%%
%%%%%%%%%%%%%%%%%%%%%%%%%%%%%%%%%%%%%%%%%%%%%%%%%%%%%%

%%%%%%%%%%%%%%%%%%%%%%%%%%%
%%%  Sec. VIII
%%%%%%%%%%%%%%%%%%%%%%%%%%%
\section{$F(R)$ bigravity and cosmological reconstruction \label{SII}}

%%%%%%%%%%%%%%%%%%%%%%%%%%%
%%%  Sec. VIII A
%%%%%%%%%%%%%%%%%%%%%%%%%%%
\subsection{$F(R)$ bigravity}

We start with reviewing $F(R)$ bigravity proposed 
in Ref.~\cite{Nojiri:2012zu}. 
The consistent model of bimetric gravity, which includes two metric tensors
$g_{\mu\nu}$ and $f_{\mu\nu}$, was proposed in Ref.~\cite{Hassan:2011zd}.
It contains the massless spin-two field, corresponding to graviton, and
massive spin-two field.
It has been shown that the Boulware-Deser ghost~\cite{Boulware:1974sr} 
does not appear in such a theory.

We consider the following action: 
\begin{align}
\label{bimetricF(R)}
S_{F} =&M_g^2\int d^4x\sqrt{-\det g}\,R^{(g)}+M_f^2\int d^4x
\sqrt{-\det f}\,R^{(f)} 
+2m^2 M_\mathrm{eff}^2 \int d^4x\sqrt{-\det g}\sum_{n=0}^{4} \bar{\beta}_n\,
e_n \left(\sqrt{g^{-1} f} \right) \nn
& - M_g^2 \int d^4 x \sqrt{-\det g}
\left\{ \frac{3}{2} g^{\mu\nu} \partial_\mu \varphi \partial_\nu \varphi
+ V(\varphi) \right\} 
+ \int d^4 x \mathcal{L}_\mathrm{M}
\left( \e^{\varphi} g_{\mu\nu}, \Phi_\mathrm{M} \right) 
\nn& - M_f^2 \int d^4 x \sqrt{-\det f}
\left\{ \frac{3}{2} f^{\mu\nu} \partial_\mu \xi \partial_\nu \xi
+ U(\xi) \right\}\, .
\end{align}
Here, $R^{(g)}$ is the scalar curvature for $g_{\mu \nu}$, 
$R^{(f)}$ is the scalar curvature for $f_{\mu \nu}$, 
$m$ is constant mass of a massive graviton, 
$M_\mathrm{eff}$ is defined by 
$\frac{1}{M_\mathrm{eff}^2} = \frac{1}{M_g^2} + \frac{1}{M_f^2}$ 
with $M_g$ and $M_f$ constants, 
and $\bar{\beta}_j$ ($j = 0, \dots, 4$) are constants. 
Moreover, $\varphi$ and $\xi$ are scalar fields, and 
$V(\varphi)$ and $U(\xi)$ are the potential of $\varphi$ and $\xi$, 
respectively. 
Furthermore, a tensor $\sqrt{g^{-1} f}$ is defined by the square root of
$g^{\mu\rho} f_{\rho\nu}$, that is,
$\left(\sqrt{g^{-1} f}\right)^\mu_{\ \rho} \left(\sqrt{g^{-1}
f}\right)^\rho_{\ \nu} = g^{\mu\rho} f_{\rho\nu}$.
For a general tensor $X^\mu_{\ \nu}$, $e_n(X)$'s are defined by
\begin{align}
\label{ek}
& e_0(X)= 1  \, , \quad
e_1(X)= [X]  \, , \quad
e_2(X)= \tfrac{1}{2}([X]^2-[X^2])\, ,
\quad e_3(X)= \tfrac{1}{6}([X]^3-3[X][X^2]+2[X^3])
\, ,\nn
& e_4(X) =\tfrac{1}{24}([X]^4-6[X]^2[X^2]+3[X^2]^2
+8[X][X^3]-6[X^4])\, ,
\quad e_k(X) = 0 ~~\mbox{for}~ k>4 \, , 
\end{align}
where $[X]$ expresses the trace of arbitrary tensor
$X^\mu_{\ \nu}$: $[X]=X^\mu_{\ \mu}$. 
By the conformal transformations
$g_{\mu\nu} \to \e^{-\varphi} g^{\mathrm{J}}_{\mu\nu}$ and
$f_{\mu\nu}\to \e^{-\xi} f^{\mathrm{J}}_{\mu\nu}$,
the action (\ref{bimetricF(R)}) is transformed as
\begin{align}
\label{FF1}
S_{F} =& M_f^2\int d^4x\sqrt{-\det f^{\mathrm{J}}}\,
\left\{ \e^{-\xi} R^{\mathrm{J}(f)} -  \e^{-2\xi} U(\xi) \right\} 
+2m^2 M_\mathrm{eff}^2 \int d^4x\sqrt{-\det g^{\mathrm{J}}}\sum_{n=0}^{4}
\bar{\beta}_n
\e^{\left(\frac{n}{2} -2 \right)\varphi - \frac{n}{2}\xi} e_n
\left(\sqrt{{g^{\mathrm{J}}}^{-1} f^{\mathrm{J}}} \right) \nn
& + M_g^2 \int d^4 x \sqrt{-\det g^{\mathrm{J}}}
\left\{ \e^{-\varphi} R^{\mathrm{J}(g)} - \e^{-2\varphi} V(\varphi) \right\}
+ \int d^4 x \mathcal{L}_\mathrm{M}
\left( g^{\mathrm{J}}_{\mu\nu}, \Phi_\mathrm{M} \right)\, .
\end{align}
Note that the kinetic terms for $\varphi$ and $\xi$ vanish. By the variations
with respect to $\varphi$ and $\xi$ as in the case of convenient $F(R)$
gravity~\cite{Nojiri:2003ft}, we obtain
\begin{align}
\label{FF2}
0 =& 2m^2 M_\mathrm{eff}^2 \sum_{n=0}^{4} 
\bar{\beta}_n \left(\frac{n}{2} -2 \right)
\e^{\left(\frac{n}{2} -2 \right)\varphi - \frac{n}{2}\xi} e_n
\left(\sqrt{{g^{\mathrm{J}}}^{-1} f^{\mathrm{J}}}\right)
+ M_g^2 \left\{ - \e^{-\varphi} R^{\mathrm{J}(g)} + 2  \e^{-2\varphi}
V(\varphi)
+ \e^{-2\varphi} V'(\varphi) \right\}\, ,\\
\label{FF3}
0 =& - 2m^2 M_\mathrm{eff}^2 \sum_{n=0}^{4} \frac{\bar{\beta}_n n}{2}
\e^{\left(\frac{n}{2} -2 \right)\varphi - \frac{n}{2}\xi} e_n
\left(\sqrt{{g^{\mathrm{J}}}^{-1} f^{\mathrm{J}}}\right)
+ M_f^2 \left\{ - \e^{-\xi} R^{\mathrm{J}(f)} + 2  \e^{-2\xi} U(\xi)
+ \e^{-2\xi} U'(\xi) \right\}\, .
\end{align}
These Eqs.~(\ref{FF2}) and (\ref{FF3}) can be solved algebraically
with respect to $\varphi$ and $\xi$ as
$\varphi = \varphi \left( R^{\mathrm{J}(g)}, R^{\mathrm{J}(f)},
e_n \left(\sqrt{{g^{\mathrm{J}}}^{-1} f^{\mathrm{J}}}\right)
\right)$ and
$\xi = \xi \left( R^{\mathrm{J}(g)}, R^{\mathrm{J}(f)},
e_n \left(\sqrt{{g^{\mathrm{J}}}^{-1}
f^{\mathrm{J}}}\right) \right)$.
Substituting the expressions of $\varphi$ and $\xi$ into (\ref{FF1}),
we acquire the action of $F(R)$ bigravity. 
We should mention, however, that it is difficult to solve Eqs.~(\ref{FF2}) and
(\ref{FF3}) with respect to $\varphi$ and $\xi$ explicitly.
Therefore, it might be easier to define the model 
in terms of the auxiliary scalars $\varphi$ and $\xi$ as in (\ref{FF1}).

We now explore the cosmological reconstruction program 
following Ref.~\cite{Nojiri:2012zu} but in a slightly extended form. 
For simplicity, we start from the minimal case: $\bar{\beta}_0=3$, $\bar{\beta}_1=-1$, $\bar{\beta}_2=\bar{\beta}_3=0$, and $\bar{\beta}_4=24$. 
In order to evaluate $\delta \sqrt{g^{-1} f}$, we examine 
two matrices $M$ and $N$, which satisfy the relation $M^2=N$. 
Since $\delta M M + M \delta M = \delta N$, we find 
$\tr \delta M = \frac{1}{2} \tr \left( M^{-1} \delta N \right)$. 
For a while, we investigate the Einstein frame action (\ref{bimetricF(R)}) 
in the minimal case and we neglect the contributions from matters. 
By the variation with respect to $g_{\mu\nu}$, we have 
\begin{align}
\label{Fbi8}
0 =& M_g^2 \left( \frac{1}{2} g_{\mu\nu} R^{(g)} - R^{(g)}_{\mu\nu} \right)
+ m^2 M_\mathrm{eff}^2 \left\{ g_{\mu\nu} \left( 3 - \tr \sqrt{g^{-1} f}
\right)
+ \frac{1}{2} f_{\mu\rho} \left( \sqrt{ g^{-1} f } \right)^{-1\, \rho}_{\qquad 
\nu}
+ \frac{1}{2} f_{\nu\rho} \left( \sqrt{ g^{-1} f } \right)^{-1\, \rho}_{\qquad 
\mu}
\right\} \nn
& + M_g^2 \left[ \frac{1}{2} \left( \frac{3}{2} g^{\rho\sigma} \partial_\rho
\varphi \partial_\sigma \varphi
+ V (\varphi) \right) g_{\mu\nu} - \frac{3}{2}
\partial_\mu \varphi \partial_\nu \varphi \right] \, .
\end{align}
On the other hand, by the variation with respect to $f_{\mu\nu}$, we find 
\begin{align}
\label{Fbi9}
0 =& M_f^2 \left( \frac{1}{2} f_{\mu\nu} R^{(f)} - R^{(f)}_{\mu\nu} \right) 
+ m^2 M_\mathrm{eff}^2 \sqrt{ \det \left(f^{-1}g\right) } 
\left \{  - \frac{1}{2}f_{\mu\rho} \left( \sqrt{g^{-1} f} 
\right)^{\rho}_{\ \nu}   - \frac{1}{2}f_{\nu\rho} 
\left( \sqrt{g^{-1} f} \right)^{\rho}_{\ \mu} 
+ \det \left( \sqrt{g^{-1} f} \right) f_{\mu\nu} \right\} \nn
& + M_f^2 \left[ \frac{1}{2} \left( \frac{3}{2} f^{\rho\sigma} \partial_\rho
\xi \partial_\sigma \xi
+ U (\xi) \right) f_{\mu\nu} - \frac{3}{2} \partial_\mu \xi \partial_\nu \xi
\right] \, .
\end{align}
We should note that $\det \sqrt{g} \det \sqrt{g^{-1} f } \neq \sqrt{f}$ in general. 
The variations of the scalar fields $\varphi$ and $\xi$ are given by
\be
\label{scalareq}
0 = - 3 \Box_g \varphi + V' (\varphi) \, ,\quad
0 = - 3 \Box_f \xi + U' (\xi) \, .
\ee
Here, $\Box_g$ ($\Box_f$) is the d'Alembertian with respect to the metric $g$ ($f$), and the prime means the derivative of the potential 
in terms of the argument as 
$V'(\varphi) \equiv \partial V (\varphi) /\partial \varphi$ 
and $U'(\xi) \equiv \partial U (\xi) /\partial \xi$. 
By multiplying the covariant derivative $\nabla_g^\mu$ with respect to the 
metric $g$ with Eq.~(\ref{Fbi8}) and using the Bianchi identity
$0=\nabla_g^\mu\left( \frac{1}{2} g_{\mu\nu} R^{(g)} - R^{(g)}_{\mu\nu} 
\right)$ and Eq.~(\ref{scalareq}), we obtain
\be
\label{identity1}
0 = - g_{\mu\nu} \nabla_g^\mu \left( \tr \sqrt{g^{-1} f} \right)
+ \frac{1}{2} \nabla_g^\mu \left\{ f_{\mu\rho} \left( \sqrt{ g^{-1} f } 
\right)^{-1\, \rho}_{\qquad \nu}
+ f_{\nu\rho} \left( \sqrt{ g^{-1} f } \right)^{-1\, \rho}_{\qquad \mu} 
\right\} \, .
\ee
Similarly, by using the covariant derivative $\nabla_f^\mu$ with respect to 
the metric $f$, from (\ref{Fbi9}) we find
\be
\label{identity2}
0 = \nabla_f^\mu \left[
\sqrt{ \det \left(f^{-1}g\right) } \left \{ - \frac{1}{2}
\left( \sqrt{g^{-1} f} \right)^{ -1 \nu}_{\ \ \ \ \ \sigma} 
g^{\sigma\mu} - \frac{1}{2}
\left( \sqrt{g^{-1} f} \right)^{ -1 \mu}_{\ \ \ \ \sigma} g^{\sigma\nu}
+ \det \left( \sqrt{g^{-1} f} \right) f^{\mu\nu} \right\} \right]\, .
\ee
In case of the Einstein gravity, the conservation law of the energy-momentum
tensor corresponds to the Bianchi identity.
In case of bigravity, however, the conservation laws of the energy-momentum 
tensor of the scalar fields are independent of the Einstein equation. 
The Bianchi identities give Eqs.~(\ref{identity1}) and (\ref{identity2}) 
independent of the Einstein equation.

We assume the FLRW universes for the metrics $g_{\mu\nu}$ and $f_{\mu\nu}$ 
and use the conformal time $t$ for the universe with the metric $g_{\mu\nu}$:
\be
\label{Fbi10}
ds_g^2 = \sum_{\mu,\nu=0}^3 g_{\mu\nu} dx^\mu dx^\nu
= a(t)^2 \left[ - dt^2 + \sum_{i=1}^3 \left( dx^i \right)^2\right] \, ,\quad
ds_f^2 = \sum_{\mu,\nu=0}^3 f_{\mu\nu} dx^\mu dx^\nu
= - c(t)^2 dt^2 + b(t)^2 \sum_{i=1}^3 \left( dx^i \right)^2 \, .
\ee
Then, $(t,t)$ and $(i,j)$ components of (\ref{Fbi8}) lead to 
\begin{align}
\label{Fbi11}
0 =& - 3 M_g^2 H^2 - 3 m^2 M_\mathrm{eff}^2
\left( a^2 - ab \right) + \left(
\frac{3}{4} {\dot\varphi}^2 + \frac{1}{2} V (\varphi) a(t)^2 \right) M_g^2 \, ,\\
\label{Fbi12}
0 =& M_g^2 \left( 2 \dot H + H^2 \right)
+  m^2 M_\mathrm{eff}^2 \left( 3a^2 - 2ab - ac \right) + \left(
\frac{3}{4} {\dot\varphi}^2 - \frac{1}{2} V (\varphi) a(t)^2 \right) M_g^2 
\, .
\end{align}
Here, $H=\dot a / a$ is the Hubble parameter as defined in Sec.~II.
On the other hand, $(t,t)$ and $(i,j)$ components of (\ref{Fbi9}) yield
\begin{align}
\label{Fbi13}
0 =& - 3 M_f^2 K^2 +  m^2 M_\mathrm{eff}^2 c^2
\left ( 1 - \frac{a^3}{b^3} \right )
+ \left( \frac{3}{4} {\dot\xi}^2 - \frac{1}{2} U (\xi) c(t)^2 \right) M_f^2 \, ,\\
\label{Fbi14}
0 =& M_f^2 \left( 2 \dot K + 3 K^2 - 2 LK \right)
+  m^2 M_\mathrm{eff}^2 \left( \frac{a^3c}{b^2} - c^2 \right)
+ \left( \frac{3}{4} {\dot\xi}^2 -
\frac{1}{2} U (\xi) c(t)^2 \right) M_f^2 \, ,
\end{align}
where $K =\dot b / b$ and $L= \dot c / c$.
Both Eqs.~(\ref{identity1}) and (\ref{identity2}) present the identical 
equation:
\be
\label{identity3}
cH = bK\ \mbox{or}\
\frac{c\dot a}{a} = \dot b\, .
\ee
If $\dot a \neq 0$, we have $c= a\dot b / \dot a$.
On the other hand, if $\dot a = 0$, we find $\dot b=0$, that is, $a$ and $b$ 
are constant and $c$ can be arbitrary. 

We redefine scalars as $\varphi=\varphi(\eta)$ and
$\xi = \xi (\zeta)$ so that we can identify $\eta$ and $\zeta$ 
with the conformal time $t$, i.e., $\eta=\zeta=t$.
Hence, we acquire 
\begin{align}
\label{Fbi19}
\omega(t) M_g^2 =& -4M_g^2 \left ( \dot{H}-H^2  \right )-2m^2 
M^2_\mathrm{eff}(ab-ac) \, , \\
\label{Fbi20}
\tilde V (t) a(t)^2 M_g^2 =&
M_g^2 \left (2 \dot{H}+4 H^2 \right ) +m^2 M^2_\mathrm{eff}(6a^2-5ab-ac)
\, , \\
\label{Fbi21}
\sigma(t) M_f^2 =& - 4 M_f^2 \left ( \dot{K} - LK  \right ) - 2m^2 
M_\mathrm{eff}^2 \left ( - \frac{c}{b} + 1 \right ) \frac{a^3c}{b^2}
\, , \\
\label{Fbi22}
\tilde U (t) c(t)^2 M_f^2 =&
M_f^2 \left ( 2 \dot{K} + 6 K^2 -2 L K  \right )
+ m^2 M_\mathrm{eff}^2 \left( \frac{a^3c}{b^2} - 2 c^2 + \frac{a^3c^2}{b^3} 
\right) \, , 
\end{align}
with
\be
\label{Fbi23}
\omega(\eta) = 3 \varphi'(\eta)^2 \, ,\quad
\tilde V(\eta) = V\left( \varphi\left(\eta\right) \right)\, ,\quad
\sigma(\zeta) = 3 \xi'(\zeta)^2 \, ,\quad
\tilde U(\zeta) = U \left( \xi \left(\zeta\right) \right) \, . 
\ee
Here, $\varphi'(\eta) \equiv \partial \varphi (\eta) /\partial \eta$ 
and $\xi'(\zeta) \equiv \partial \xi (\zeta) /\partial \zeta$. 
Thus, for arbitrary $a(t)$, $b(t)$, and $c(t)$, if we choose $\omega(t)$, 
$\tilde V(t)$, $\sigma(t)$, and $\tilde U(t)$
to satisfy Eqs.~(\ref{Fbi19})--(\ref{Fbi22}), the cosmological model 
with given evolutions of $a(t)$, $b(t)$, and $c(t)$ can be reconstructed.

%%%%%%%%%%%%%%%%%%%%%%%%%%%
%%%  Sec. VIII B
%%%%%%%%%%%%%%%%%%%%%%%%%%%
\subsection{Cosmological bouncing models \label{SIV}}

Next, we construct cosmological bouncing models. 
The physical metric, where the scalar does not directly 
couple with matter, is given by multiplying the scalar field to the metric in
the Einstein frame in (\ref{bimetricF(R)}):
$g^\mathrm{J}_{\mu\nu} = \e^{\varphi} g_{\mu\nu}$.
In the bigravity model, there appears another (reference) metric tensor
$f_{\mu\nu}$ besides $g_{\mu\nu}$.
In our model, since the matter only couples with $g_{\mu\nu}$, the physical
metric could be given by $g^\mathrm{J}_{\mu\nu}$.

In our formulation, it is convenient to use the conformal time description.
The conformally flat FLRW universe metric is given by
\be
\label{Fbi31}
ds^2 = \tilde a(t)^2 \left[ - dt^2 + \sum_{i=1}^3 \left( dx^i \right)^2
\right] \, .
\ee
This equation~(\ref{Fbi31}) with $g^\mathrm{J}_{\mu\nu} = \e^{\varphi} g_{\mu\nu}$ shows $\e^{\varphi(t)} a(t)^2 = \tilde a(t)^2$, that is, 
$\varphi = - 2 \ln a(t) + \ln \tilde a(t)$. 
By using (\ref{Fbi23}), we find
\be
\label{FFbi3}
\omega(t) = 12 \left( H - \tilde H \right)^2 \, .
\ee
Here, $\tilde H \equiv \frac{1}{\tilde a}\frac{d\tilde a}{dt}$.

In the following, by making the choice $a(t)=b(t)=1$, we explicitly construct
the model generating the bouncing behavior. 
We should remark that the choice $a(t)=b(t)=1$ satisfies the constraint 
(\ref{identity3}).

When $a(t)=b(t)=1$, the Einstein frame metric $g_{\mu\nu}$
expresses the flat Minkowski space, although the metric we observe is given by
$g^\mathrm{J}_{\mu\nu}$. 
Equations~(\ref{Fbi19}), (\ref{Fbi20}), (\ref{Fbi21}), and (\ref{Fbi22}) with
(\ref{FFbi3}) are simplified as follows
\begin{align}
\label{Fbi19C}
\omega (t) M_g^2 =& 12 M_g^2 \tilde H^2 = m^2 M_\mathrm{eff}^2
\left( c - 1\right) \, , \\
\label{Fbi20C}
\tilde V (t) M_g^2 =& m^2 M_\mathrm{eff}^2 \left( 1 - c \right)
= - 6 M_g^2 \tilde H^2 \, , \\
\label{Fbi21C}
\sigma(t) M_f^2 =& 2 m^2 M_\mathrm{eff}^2 \left( c - 1 \right)
= 12 M_g^2 \tilde H^2 \, , \\
\label{Fbi22C}
\tilde U (t) M_f^2 =& m^2 M_\mathrm{eff}^2 c \left( 1 - c \right) 
= - 6 M_g^2 \tilde H^2
\left( 1 + \frac{ 6 \tilde H^2}{m^2 M_\mathrm{eff}^2} \right) \, .
\end{align}
Equation~(\ref{Fbi19C}) can be solved with respect to $c$ as
\be
\label{FFFbi1}
c = 1 + \frac{ 6 \tilde H^2}{m^2 M_\mathrm{eff}^2} \, .
\ee
We should note that both $\omega(t)$ and $\sigma(t)$ are positive 
and hence there does not appear any ghost in the theory.

We now study the bouncing solution
\be
\label{FbiB1}
\tilde a (t) \sim \e^{\bar{\alpha} t^2}\, , 
\ee
with $\bar{\alpha}$ a positive constant. 
Since 
\be
\label{FbiB2}
\tilde H \sim 2 \bar{\alpha} t\, ,
\ee
we find 
\be
\label{FbiB3}
c (t) = 1 + \frac{12 \bar{\alpha}^2 M_g^2 t^2}{m^2 M_\mathrm{eff}^2}\, ,
\ee
and
\be
\label{FbiB4}
\omega (\eta) = 12 \bar{\alpha} M_g^2 \eta^2\, ,\quad 
\tilde V (\eta) = - 12 \bar{\alpha}^2 \eta^2\, ,\quad 
\sigma (\zeta) = \frac{24 \bar{\alpha}^2 M_g^2 \zeta^2}{M_f^2}\, ,\quad 
\tilde U (\eta) = - \frac{12 \bar{\alpha}^2 M_g^2 \zeta^2}{M_f^2} 
\left( 1 + \frac{12 \bar{\alpha}^2 M_g^2 \zeta^2}{m^2 M_\mathrm{eff}^2} \right)\, .
\ee

Consequently, for Eq.~(\ref{FbiB1}), 
the solutions in (\ref{FbiB4}) can be obtained. 
Moreover, the exponential form of the scale factor in Eq.~(\ref{FbiB1}) 
is equivalent to that in Eq.~(\ref{b1}), which can lead to 
the bouncing behavior. 
This means that in the flat FLRW universe, 
for an exponential form of the scale factor in $F(R)$ bigravity, 
bounce cosmology can be realized, similarly to that in $F(R)$ gravity, 
as demonstrated in Sec.~III A. 
For the case that the scale factor has an exponential form 
in Eq.~(\ref{FbiB1}), 
%or (\ref{FbiB1}), 
in terms of the physical metric, the bouncing behavior in $F(R)$ bigravity 
is the same as that in $F(R)$ gravity. 
On the other hand, for this case, in the reference metric, i.e., 
the fiducial metric existing only in $F(R)$ bigravity, 
it is clearly seen from Eqs.~(\ref{FbiB1}) and (\ref{FbiB3}) that 
also in this reference metric, the bouncing behavior can occur, 
but the contraction and expansion rates are different each other. 
In the physical metric, 
$\tilde{H} = \dot{\tilde{a}}/\tilde{a} \sim 2 \bar{\alpha} t$ as given 
by Eq.~(\ref{FbiB2}), while in the reference metric, 
$\dot{c}/c = 2 \mathcal{I} t/\left(1 + \mathcal{I} t^2 \right)$ 
with $\mathcal{I} \equiv 12 \bar{\alpha}^2 M_g^2 /\left( m^2 M_\mathrm{eff}^2\right)$. The ratio of $\tilde{H}$ to $\dot{c}/c$ reads 
$\mathcal{R} \equiv \tilde{H}/\left(\dot{c}/c\right) \simeq \bar{\alpha} \mathcal{I}^{-1} 
\left(1+ \mathcal{I} t^2 \right)$. Thus, when $\mathcal{I} t^2 \gg 1$, 
the contraction and expansion rates in the physical metric 
are much larger than those in the reference metric, 
while for $\mathcal{I} t^2 = \mathcal{O}(1)$, 
namely, around the bouncing epoch, 
the ratio defined above 
becomes $\mathcal{R} \sim m^2 M_\mathrm{eff}^2/\left(\bar{\alpha}^2 M_g^2 \right)$. This implies that whether the contraction and expansion rates 
in the physical metric is larger or smaller than those in the reference metric depends on the model parameters. 

%%%%%%%%
Furthermore, since the form of the scale factor $\tilde{a}(t)$ 
in Eq.~(\ref{FbiB1}) in the physical metric is equivalent to that 
of $a(t)$ in Eq.~(\ref{b1}), it is considered that 
the same consequences as in Sec.~III A 
in terms of the cosmological evolution and values of 
$F'(R)$~\cite{Nariai:1973eg, Gurovich:1979xg} and $F''(R)$ would be obtained. 
%%%%%%%%

%%%%%%%%%%%%%%%%%%%
%%%  Sec. IX
%%%%%%%%%%%%%%%%%%%
\section{Conclusions}

In the present paper, we have reconstructed $F(R)$ gravity models where bounce cosmology can occur. As concrete models, 
we have demonstrated the cases 
that in the flat FLRW universe, the scale factor has exponential and power-law forms in Eqs.~(\ref{b1}) and (\ref{eq:3.9}), respectively. 
For an exponential form of the scale factor in Eq.~(\ref{b1}), 
an $F(R)$ gravity model with 
the second order polynomial in terms of $R$ is reconstructed, 
whereas for the power-law form, the resultant $F(R)$ function is proportional 
to $R$, equivalent to that in general relativity. 
In addition, 
we have investigated the perturbations from the background solutions and 
examined the explicit stability conditions for these reconstructed models. 
As a result, it has been found that these models could be stable 
because the stability conditions can be satisfied. 
It has to be stressed that the matter bounce scenario~\cite{B-R-BCS, Cai:2013kja} (for a specific case) proposed by Brandenberger et al. is able to be reproduced also in $F(R)$ gravity. 

Also, we have explored a sum of exponentials 
form of the scale factor in Eq.~(\ref{eq:FR7-6-VA-01-1}) 
in order to derive an $F(R)$ gravity model 
in which the bounce 
in the early universe and the late-time accelerated expansion of 
the universe can be realized in a unified manner. 
In this case, a second order polynomial $F(R)$ gravity model 
is derived as in a model where the scale factor consists of 
a single exponential term. 
For this model, 
we have analyzed the stability condition and 
confirmed that it can be met. 
Accordingly, it is considered that the model with the sum of exponentials 
form of the scale factor could be stable. 
It is remarkable that the $R^2$-gravity theory of the same type as the one realizing inflation occurs as the theory which gives rise to bounce cosmology does. 

Furthermore, 
in the FLRW universe with non-zero spatial curvature, 
for the scale factor with an exponential form in Eq.~(\ref{sol}), 
we have reconstructed 
a second order polynomial $F(R)$ gravity model 
and an $F(R)$ gravity model with 
a term proportional to $R$ and that proportional to 
$1/R$~\cite{Carroll:2003wy}. 
As a consequence, 
it has been seen that only in the non-flat FLRW universe 
with non-zero spatial curvature, 
a solution can exist, and that if the cosmic curvature vanishes, 
we can obtain only the de Sitter solution and 
hence bounce cosmology cannot be realized. 

Therefore, 
when the scale factor is given by an exponential form in Eq.~(\ref{sol}), 
by using the reconstruction method, 
we have derived $F(R)$ gravity models realizing bounce cosmology. 
Regarding one model leading to bounce cosmology, 
we have also analyzed the stability conditions and 
confirmed that these conditions can be satisfied 
and thus this model can be stable. 

Moreover, we have reconstructed an $F(R)$ bigravity model in which 
bounce cosmology can be realized. 
It has been verified that in $F(R)$ bigravity, 
for an exponential form of the scale factor in Eq.~(\ref{FbiB1}), 
in the flat FLRW universe bounce cosmology can be realized. 
It is interesting to emphasize that 
not only in the physical metric but also in the reference metric 
the bouncing behavior can happen. 
Also, if the cosmic time is very far past or 
future from the bouncing epoch, 
the contraction and expansion rates in the physical metric 
are much larger than those in the reference metric. 
On the other hand, around the bouncing epoch, 
if the values of the model parameters are determined, 
we can see which contraction and expansion rates 
in the physical or reference metric are larger or smaller.

%%%%%%%%%%%%%%%%%%%%%%%%
%%%  Acknowledgments
%%%%%%%%%%%%%%%%%%%%%%%%
\section*{Acknowledgments}

S.D.O. sincerely acknowledges the Kobayashi-Maskawa Institute (KMI) 
visitor program for very kind hospitality in Nagoya University, 
where this work has greatly progressed. 
The work is supported in part 
by the JSPS Grant-in-Aid for 
Young Scientists (B) \# 25800136 (K.B.);\ 
that for Scientific Research 
(S) \# 22224003 and (C) \# 23540296 (S.N.);\ 
and 
MINECO (Spain), FIS2010-15640 and
AGAUR (Generalitat de Ca\-ta\-lu\-nya), contract 2009SGR-345, 
and MES project 2.1839.2011 (Russia) 
(S.D.O. and A.N.M.).

%%%%%%%%%%%%%%%%%%%%%%%%%%%%%%%%%
%% thebibliography environment
%%%%%%%%%%%%%%%%%%%%%%%%%%%%%%%%%

\end{document}